\documentclass[prb,aps,amssymb,twocolumn, superscriptaddress, preprintnumbers, showpacs, showkeys]{revtex4-2}
\usepackage{amsmath}
\usepackage{amsthm}
\usepackage{listings}
\usepackage{braket}
\usepackage{graphicx}
\usepackage{multirow}
\usepackage[utf8]{inputenc}
\usepackage[colorlinks, linkcolor={red!80!black},citecolor={blue!70!black},urlcolor={blue!80!black}]{hyperref}
\usepackage{psfrag}
\usepackage{comment}
\usepackage{bm}
\usepackage{xcolor}
\usepackage{lineno}

\newcommand{\beq}{\begin{equation}}
	\newcommand{\eneq}{\end{equation}}

\begin{document}
	\title{Non-Hermitian skin effect enforced by nonsymmorphic symmetries}
	
	\author{Yutaro Tanaka}
	\email{tanakay@stat.phys.titech.ac.jp}
	\affiliation{Department of Physics, Tokyo Institute of Technology, Tokyo 152-8551, Japan}
	
	\author{Ryo Takahashi}
	\affiliation{Advanced Institute for Materials Research (WPI-AIMR), Tohoku University, Sendai 980-8577, Japan}
	
	\author{Ryo Okugawa}
	\affiliation{Department of Applied Physics, Tokyo University of Science, Tokyo 125-8585, Japan}
	
	\begin{abstract}
		Crystal symmetries play an essential role in band structures of non-Hermitian Hamiltonian. In this paper, we propose a non-Hermitian skin effect (NHSE) enforced by nonsymmorphic symmetries. We show that the NHSE inevitably occurs if a two-dimensional non-Hermitian system satisfies conditions derived from the nonsymmorphic symmetry of the doubled Hermitian Hamiltonian. This NHSE occurs in symmetry classes with and without time-reversal symmetry. The NHSE enforced by nonsymmorphic symmetries always occurs simultaneously with the closing of the point gap at zero energy. We also show that such a NHSE can occur in specific three-dimensional space groups with nonsymmorphic symmetries. 
	\end{abstract}
	
	\maketitle
	
	\section{Introduction} 
	Non-Hermiticity is universal in physics because open quantum and classical systems are effectively described by non-Hermitian matrices \cite{Gamow1928, PhysRev.56.750,  FESHBACH1958357, FESHBACH1962287, Kane:2014uu, Huber:2016tg, RevModPhys.88.035002, feng2017non, el2018non, kozii2017non,  PhysRevLett.121.026403, PhysRevB.99.201107}.
	With recent experimental advances, non-Hermitian band structures have been intensively investigated.
	Intriguingly, non-Hermiticity allows unique non-Hermitian band topology without Hermitian counterparts \cite{doi:10.1080/00018732.2021.1876991, RevModPhys.93.015005}, $e.g.$, 
	the non-Hermitian skin effect (NHSE) \cite{PhysRevLett.116.133903, PhysRevB.97.121401, PhysRevX.8.031079, PhysRevLett.121.086803, PhysRevLett.121.136802, PhysRevLett.121.026808, PhysRevX.9.041015, PhysRevB.99.201103, PhysRevB.99.235112, PhysRevLett.123.016805, PhysRevLett.123.066404, PhysRevLett.123.170401, PhysRevLett.123.246801, PhysRevResearch.1.023013, PhysRevLett.124.056802, PhysRevLett.124.086801, PhysRevB.101.195147, PhysRevLett.125.126402, PhysRevResearch.2.023265,  PhysRevB.103.165123, PhysRevB.104.L161106, PhysRevB.104.165117, PhysRevB.105.075128, yokomizo2021non, PhysRevB.105.L180406, PhysRevB.105.245143, PhysRevLett.129.070401, PhysRevLett.129.086601, zhang2022universal, doi:10.1063/5.0097530, PhysRevResearch.4.033122, PhysRevX.13.021007}. 
	The NHSE is a drastic difference in energy spectrum between a periodic boundary condition (PBC) and an open boundary condition (OBC) \cite{doi:10.1080/00018732.2021.1876991, RevModPhys.93.015005}.
	In particular, owing to the NHSEs,
	a systematic understanding of non-Hermitian band topology is still elusive in high-dimensional systems.
	
	To understand the band structures, the concept of topological phases protected by crystal symmetries \cite{PhysRevLett.106.106802, PhysRevLett.108.266802, hsieh2012topological, slager_space_2013, PhysRevB.89.155114, murakami2017emergence, PhysRevLett.119.246402, schindler2018higher, PhysRevB.97.205136, PhysRevX.8.031070, fang2017rotation, PhysRevB.101.115120, PhysRevResearch.2.013300, he2020quadrupole,  PhysRevResearch.2.043274, Kim:21, PhysRevB.105.064517, PhysRevB.105.115119, PhysRevLett.128.116802, PhysRevLett.129.046802, PhysRevLett.129.056403,PhysRevB.107.245148} has been applied to non-Hermitian systems.
	Previous works have revealed a relationship between non-Hermitian energy spectra and band topology through symmorphic symmetries, including inversion \cite{PhysRevB.99.041202, PhysRevB.102.241202, PhysRevB.103.205205, PhysRevB.103.L201114, PhysRevB.104.035424, PhysRevB.105.085109}, mirror \cite{PhysRevB.104.035424, PhysRevResearch.2.022062}, and rotation \cite{PhysRevB.102.205118, PhysRevB.104.035424, PhysRevLett.128.226401, wakao2022discriminant}.    
	In contrast, little has been done to examine the effects of nonsymmorphic symmetries ($e.g.$ screw rotation and glide mirror) on non-Hermitian Hamiltonians although they add strong constraints to band structures in Hermitian systems, such as symmetry-enforced gaplessness \cite{PhysRevLett.115.126803, PhysRevB.94.155108, bzduvsek2016nodal, PhysRevB.93.155140, chen2018topological, PhysRevLett.117.096404, PhysRevB.96.155206}.
	
	In this paper, we propose a class of NHSEs enforced by nonsymmorphic symmetries.
	This NHSE necessarily occurs if a non-Hermitian Hamiltonian satisfies certain conditions of symmetry and matrix size.
    We also clarify how the conditions depend on the presence of time-reversal symmetry.
	We also show that when the nonsymmorphic-symmetry-enforced NHSE occurs, the point gap always closes at $E=0$ under the full PBC. 
	The gap closing is a non-Hermitian counterpart of symmetry-enforced gaplessness in Hermitian systems. 
    This NHSE can occur in 2D systems with and  without time-reversal symmetry.
    
    \section{Nonsymmorphic-symmetry-enforced NHSE}\label{secII}
    To begin with, we explain the point-gap topology for NHSEs \cite{PhysRevX.8.031079, PhysRevX.9.041015, PhysRevB.99.235112, PhysRevLett.124.086801}.
    A point gap at a wavevector $\bm{k}$ is open at a reference energy $E$
	if the Hamiltonian $H(\boldsymbol{k})$ satisfies ${\rm det}[H(\boldsymbol{k})-E]\neq 0$.
	When the point gap is open at $E=0$ on a line along the $k_{i}~(i=x,y,z)$ direction,
	we can define a winding number along the $k_{i}$ direction as
	\begin{equation}\label{winding}
		W_{{i}}(k_{\perp})=\int^{2\pi}_{0}\frac{d k_i}{2\pi i}\frac{\partial}{\partial k_i}\log \det H(\boldsymbol{k}),
	\end{equation}
	where $k_{\perp}$ is the wave vector perpendicular to $k_{i}$. 
	If $W_{i}(k_{\perp})$ is nonzero,
	the NHSE occurs under the OBC in the $i$ direction \cite{PhysRevX.8.031079, PhysRevX.9.041015, PhysRevLett.124.086801, zhang2022universal}.
	Throughout this paper, we focus on the reference energy $E=0$.

The winding number is related to Hermitian band topology.
To see a relationship, we introduce a doubled Hermitian Hamiltonian
\begin{align}\label{eq:extended_hermitian_2d}
	\tilde{H}(\boldsymbol{k})
	=\begin{pmatrix} 0 & H(\boldsymbol{k}) \\ H^{\dagger}(\boldsymbol{k}) & 0
	\end{pmatrix}.
\end{align}
The point-gap topology of a non-Hermitian Hamiltonian $H(\boldsymbol{k})$ corresponds to the band topology of the doubled Hermitian Hamiltonian $\tilde{H}(\boldsymbol{k})$,
which has chiral symmetry
\cite{PhysRevB.78.195125, RevModPhys.88.035005,  PhysRevX.8.031079, PhysRevX.9.041015}.
The chiral symmetry is described as $\Gamma \tilde{H}(\boldsymbol{k}) \Gamma^{-1}=-\tilde{H}(\boldsymbol{k})$
with $\Gamma={\rm diag}(\boldsymbol{1}_{N}, -\boldsymbol{1}_{N})$,
where $\boldsymbol{1}_{N}$ is the identity matrix that has the same matrix size as $H(\bm{k})$.
Doubled Hermitian Hamiltonians are also characterized by a winding number equivalent to Eq.(\ref{winding}).
Therefore, we can diagnose whether the winding number $W_{i}(k_{\perp})$ is nonzero from information on the topology of the doubled Hermitian Hamiltonian (see Appendix \ref{appendix:review_doubled_winding} for details of the connection between the topology of $\tilde{H}(\boldsymbol{k})$ and $W_{i}(k_{\perp})$).

    In this section, we first consider a 2D non-Hermitian Hamiltonian in 
    class AII$^{\dagger}$  with time-reversal symmetry 
    $
	\mathcal{T}H^{T}(\boldsymbol{k})\mathcal{T}^{-1}=H(-\boldsymbol{k}) 
	$, where $\mathcal{T}$ is a unitary matrix satisfying $\mathcal{T}\mathcal{T}^{*}=-1$ \cite{PhysRevX.9.041015}. 
    We demonstrate the conditions under which a NHSE unavoidably occurs in the presence of the time-reversal symmetry.
    Initially, we assume that the matrix size of the Hamiltonian is $N=4M+2$ with a nonnegative integer $M$.
    Additionally, we suppose that the Hamiltonian satisfies
    \begin{align}\label{eq:generalized_screw_symmetry}
		U_{g}^{\boldsymbol{k}}H(\boldsymbol{k})=\Bigl(U_{g}^{C_{2x}\boldsymbol{k}}
		H(C_{2x}\boldsymbol{k}) \Bigr)^{\dagger},
	\end{align}
	where $C_{2x}$ represents the two-fold rotation around the $x$ axis, i.e. $C_{2x}\boldsymbol{k}=(k_{x}, -k_{y})$.
	Here $U_{g}^{\boldsymbol{k}}$ is a unitary matrix defined as 
	\begin{align}\label{eq:generalized_screw_symmetry2}
		U_{g}^{\boldsymbol{k}}:=c_{s}e^{i\frac{k_x}{2}}A^{\boldsymbol{k}}_{g},
	\end{align}
	where $c_s=i(1)$ for $C_{2x}^2=-1(+1)$ in spinful (spinless) systems,
	and $A^{\boldsymbol{k}}_{g}$ is an arbitrary $(4M+2) \times (4M+2)$ unitary matrix that is periodic in the Brillouin zone.
     In addition, in the presence of time-reversal symmetry, $U_{g}^{\boldsymbol{k}}$ needs to satisfy a relationship with $\mathcal{T}$
    \begin{align}\label{eq:classAIIdagger_trs_ag}
 \mathcal{T}(U_g^{\boldsymbol{k}})^{*}=-(U_g^{-C_{2x}\boldsymbol{k}})^{\dagger}\mathcal{T}.
\end{align}
	As we discuss below, Eq.~(\ref{eq:generalized_screw_symmetry}) is associated with a screw rotation $g=\{ C_{2x} | \hat{\boldsymbol{x}}/2 \}$ symmetry,
	which is a twofold rotation $C_{2x}$ followed by translation by a half of a lattice vector $\hat{\boldsymbol{x}}$.
    In this paper, we use Seitz notation $g= \{p_{g}| \boldsymbol{t}_{g}\}$ for a spatial transformation $g$, 
    which transform a point $\boldsymbol{r}$ as $\boldsymbol{r} \rightarrow p_{g}\boldsymbol{r} +\boldsymbol{t}_{g}$.
    Here $p_{g}$ is an element of the point group, and $\boldsymbol{t}_{g}$ represents a translation.
    
	 Because the screw symmetry enforces NHSEs owing to the above condition, 
	we refer to such NHSEs as nonsymmorphic-symmetry-enforced NHSEs.
	At the same time, the condition leads to the closing of the point gap at $E=0$ on both $k_y=0$ and $k_y=\pi$ lines under the full PBC.

\subsection{2D Tight-binding model in class AII$^{\dagger}$}
\begin{figure}
		\includegraphics[width=1.\columnwidth]{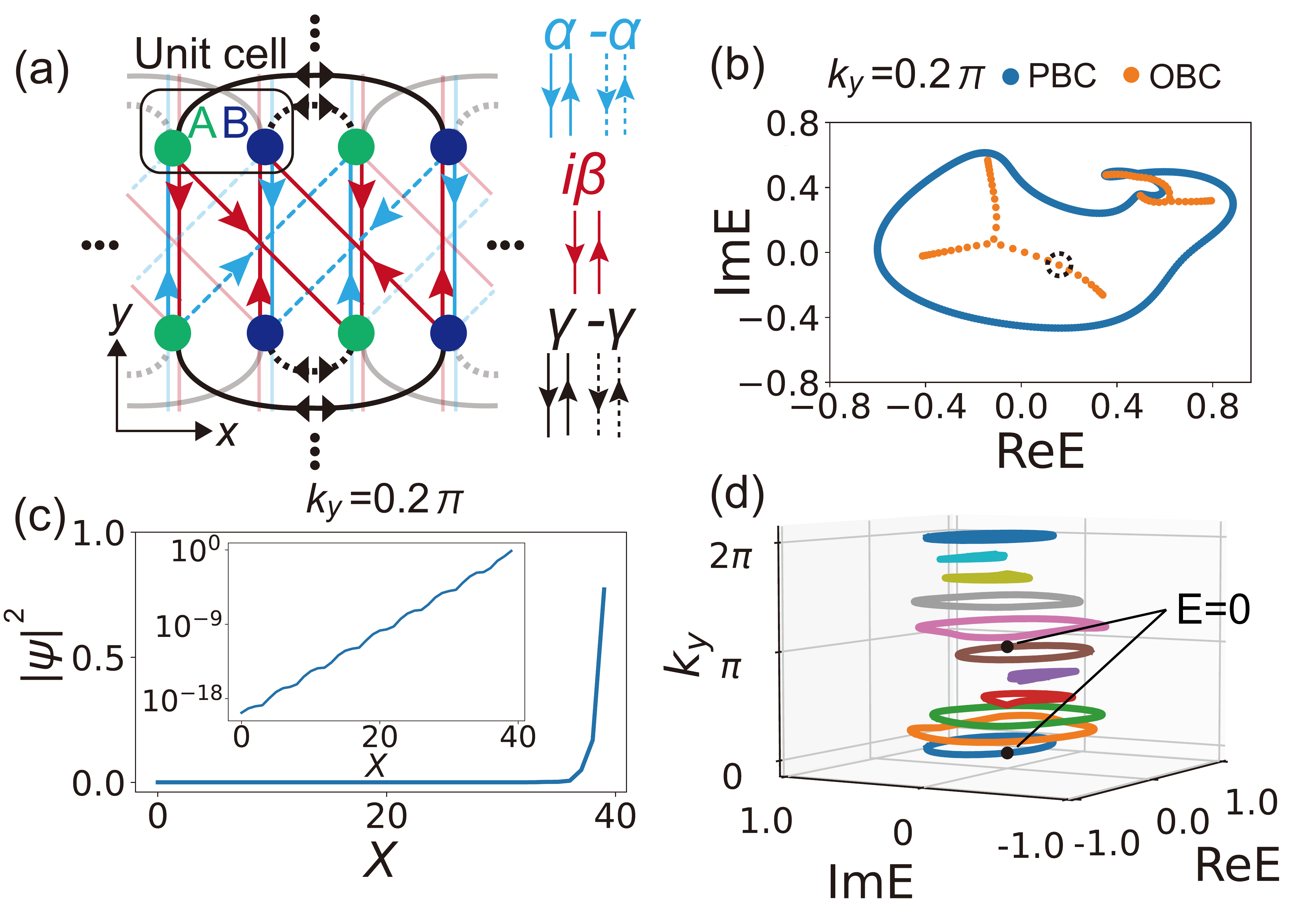}
\caption{Energy spectra of $H_{\rm 2D}(\boldsymbol{k})$ with the parameters $\alpha=\beta=0.3$ and $\gamma=0.2$. We take the PBC in the $y$ direction for all the calculations. (a) Schematic diagram of our model in the real space. The rounded rectangle  indicates the unit cell.  (b) The energy spectra on the $k_y=0.2\pi$ line under the PBC and the OBC in the $x$ direction. (c) The spatial distributions of the skin mode in the dotted circle in (b). 
The inset shows the spatial distribution in log scale. The system size in the $x$ direction is 40. 
(d) The energy spectra in $E$-$k_y$ space under the full PBC. 
}\label{fig:classAIIdagger_model}
	\end{figure}

For concreteness, we introduce a 2D two-band tight-binding model with the two degrees of freedom $A$ and $B$ in each unit cell on a square lattice [Fig.~\ref{fig:classAIIdagger_model}(a)] in class AII$^{\dagger}$: 
\begin{align}
	H_{\rm 2D}=H_{\alpha}+H_{\beta}+H_{\gamma}.
\end{align}
Here, $H_{\alpha}$ and $H_{\beta}$ are the non-Hermitian parts written as
\begin{align}
	H_{\alpha}=\alpha \sum_{\boldsymbol{R}} \sum_{\boldsymbol{r}_{x}=0, \boldsymbol{e}_{x}}&(-1)^{\boldsymbol{r}_x \cdot \boldsymbol{e}_x}  \biggl( 
	\ket{ \boldsymbol{R}+\boldsymbol{r}_{x}+\boldsymbol{e}_y, A}\bra{\boldsymbol{R},A}\nonumber \\
	&+\ket{\boldsymbol{R}-\boldsymbol{r}_{x}-\boldsymbol{e}_{y},B}\bra{\boldsymbol{R},B}\biggr), \nonumber \\
	H_{\beta}=i\beta \sum_{\boldsymbol{R}} \sum_{\boldsymbol{r}_{x}=0, \boldsymbol{e}_{x}}
 & \biggl( 
	\ket{ \boldsymbol{R}+\boldsymbol{r}_{x}-\boldsymbol{e}_y, A}\bra{\boldsymbol{R},A}\nonumber \\
	&+\ket{\boldsymbol{R}-\boldsymbol{r}_{x}+\boldsymbol{e}_{y},B}\bra{\boldsymbol{R},B}\biggr), 
\end{align}
 and 
 $H_{\gamma}$ is the Hermitian part written as
\begin{align}
H_{\gamma} = \gamma \sum_{\boldsymbol{R}} \sum_{\boldsymbol{r}_{x}=0, \boldsymbol{e}_{x}}
  & \biggl(
  -\ket{ \boldsymbol{R}+\boldsymbol{e}_x, A}\bra{\boldsymbol{R},B}\nonumber \\
 & +\ket{ \boldsymbol{R}-\boldsymbol{e}_x, A}\bra{\boldsymbol{R},B}+{\rm h.c.}  \biggr),
\end{align}
where $\boldsymbol{R}=(x,y)$ is the position of the unit cell, and $\boldsymbol{e}_{x}$ and $\boldsymbol{e}_{y}$ are the unit vectors in the $x$ and $y$ directions, respectively.

Let us introduce the Fourier transformations
$
	\ket{\boldsymbol{k}}=\frac{1}{\sqrt{N'}}\sum_{\boldsymbol{R}}e^{i\boldsymbol{k}\cdot \boldsymbol{R}}\ket{\boldsymbol{R}},
$
where $N'$ is the number of unit cells.
By using the Fourier transformation under the PBC, we obtain the Bloch Hamiltonian 
    \begin{equation}
        H_{\rm 2D}(\boldsymbol{k})=\begin{pmatrix}
        d(\boldsymbol{k}) & 2i\gamma \sin k_x \\
        -2i\gamma \sin k_x & d(-\boldsymbol{k}) \\
        \end{pmatrix},
    \end{equation}
    where $d(\boldsymbol{k})= \alpha(1-e^{-i k_x})e^{-ik_y}+i\beta(1+e^{-ik_x})e^{ik_y}$.
    The Hamiltonian has time-reversal symmetry: $\sigma_y (H_{\rm 2D}(\boldsymbol{k}))^{T}(\boldsymbol{k})\sigma_y^{-1} =H_{\rm 2D}(-\boldsymbol{k})$, where $\sigma_y$ is the Pauli matrix. 
    The schematic diagram of this model in the real space is shown in Fig.~\ref{fig:classAIIdagger_model}(a). 
    The Hamiltonian $H_{\rm 2D}(\boldsymbol{k})$ satisfies the condition in Eq. (\ref{eq:generalized_screw_symmetry}) with $c_s=i$ and  $A^{\boldsymbol{k}}_{g}={\rm diag}(1,-e^{-ik_x})$. Figure \ref{fig:classAIIdagger_model}(b) shows energy spectra of $H_{\rm 2D}(\boldsymbol{k})$ on the $k_{y}=0.2\pi$ line both under the full PBC and under the OBC in the $x$ direction.
    The model has a nonzero winding number in Eq.~(\ref{winding}) under the full PBC.
	Figure \ref{fig:classAIIdagger_model}(c) shows that the eigenstate on the $k_y = 0.2\pi$ line is localized at the right edge. 
	These results indicate that a NHSE occurs in the model.  
	Furthermore, the point gap closes at $E=0$ on the $k_y=0$ and $k_y=\pi$ lines under the full PBC [Fig.~\ref{fig:classAIIdagger_model}(d)]. 

\begin{figure}
	\includegraphics[width=1.\columnwidth]{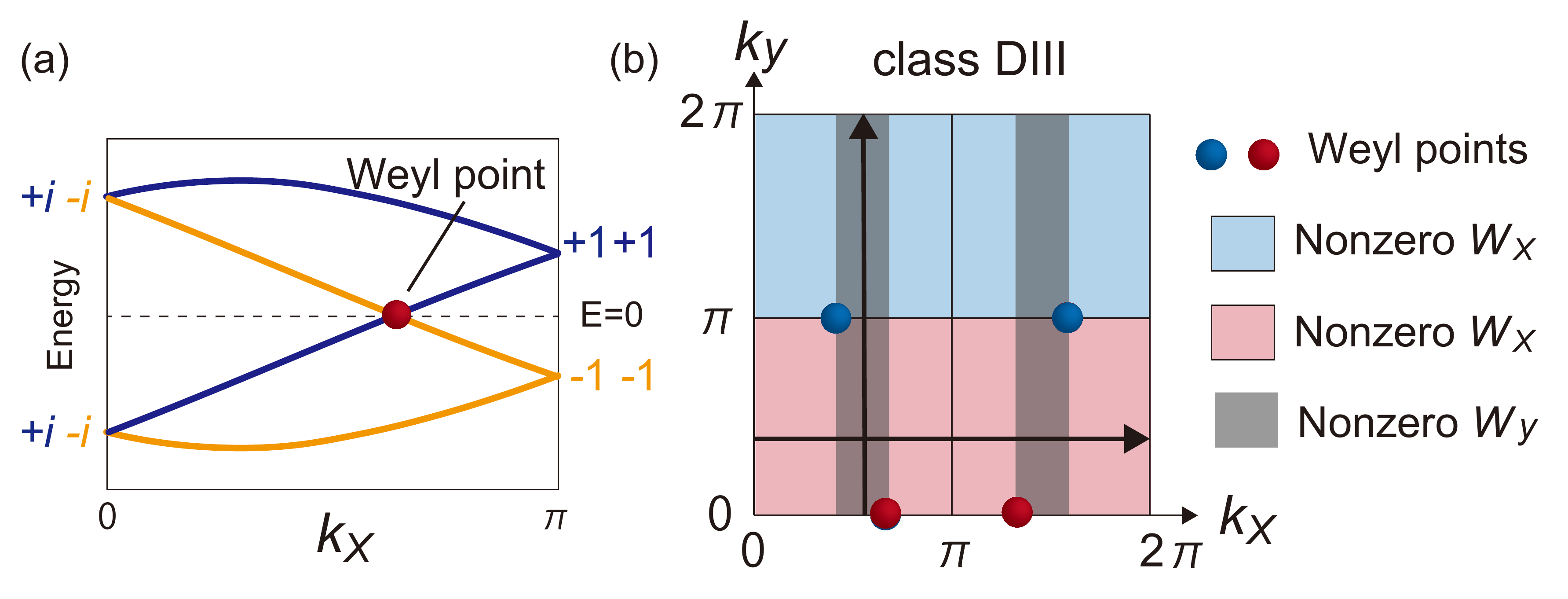}
	\caption{
		(a) Schematic band structures of a doubled Hermitian Hamiltonian with  $g = \{ C_{2x} | \hat{\boldsymbol{x}}/2 \}$ symmetry and time-reversal symmetry along the screw-invariant line. $\Gamma$ and $\tilde{U}^{\boldsymbol{k}}_{g}$ satisfy anticommutation relation. Colors of the bands indicate the opposite sign of the screw eigenvalues.
			(b) An illustrative example of the positions of the Weyl points in class DIII with $c_g=-1$ and the regions with the nonzero winding number in the Brillouin zone.
			The arrows indicate the direction of integration of $W_{i} (k_{\perp})$. The colors (red and blue) of the points represent the opposite chirality of the Weyl points. The red and blue colors of the nonzero $W_x(k_y)$ regions indicate the opposite sign of the winding number: $W_x(k_y)=-W_x(-k_y)$.
	} \label{fig:concept_AIIdagger}
\end{figure}

Here we consider a doubled Hermitian Hamiltonian of $H_{\rm 2D}(\bm{k})$ to find the band topology that leads to NHSE.
The doubled Hermitian Hamiltonian has a screw symmetry written as 
$\tilde{U}^{\boldsymbol{k}}_{\{ C_{2x} | \hat{\boldsymbol{x}}/2 \}}\tilde{H}(\boldsymbol{k})\Big{(}\tilde{U}^{\boldsymbol{k}}_{\{ C_{2x} | \hat{\boldsymbol{x}}/2 \}}\Big{)}^{-1}=\tilde{H}(C_{2x}\boldsymbol{k})$
with
\begin{align}
\tilde{U}^{\boldsymbol{k}}_{\{ C_{2x} | \hat{\boldsymbol{x}}/2 \}}=
	\begin{pmatrix}
			 & -e^{-ik_x}(A_{g}^{C_2\bm{k}})^{\dagger} \\
			A^{\boldsymbol{k}}_{g} & 
	\end{pmatrix}.
\end{align}
The doubled Hermitian Hamiltonian hosts 2D Weyl points on the screw-invariant lines.
The schematic band structures are illustrated in Fig~\ref{fig:concept_AIIdagger}.
The 2D Weyl points originate from band crossings of eigenstates with different screw-symmetry eigenvalues.
In general, a winding number can be nonzero in the presence of 2D Weyl points
\cite{Matsuura_2013, RevModPhys.88.035005,  PhysRevLett.118.156401, PhysRevB.105.035429, PhysRevB.106.045126}.
The relationship between a winding number and 2D Weyl points is given in Appendix \ref{appendix:review_doubled_winding}. 
Therefore, we can see that the model shows the NHSE.
	
\subsection{Conditions for the screw symmetry}\label{sec:condition_for_NHSE}

Hereafter, we discuss general conditions to systematically realize NHSEs protected by screw symmetry, including the NHSE in the above 2D model.
First, we briefly review conditions of non-Hermitian Hamiltonians for doubled Hermitian Hamiltonians in Eq.~(\ref{eq:extended_hermitian_2d}) to have crystal symmetry $g=\{p_{g}| \boldsymbol{t}_{g}\}$ \cite{PhysRevB.104.035424}.
The symmetry for the doubled Hermitian Hamiltonian is expressed as 
\begin{align}\label{eq:extended_hermitian_symmetry}
	\tilde{U}^{\boldsymbol{k}}_{g}\tilde{H}(\boldsymbol{k})\Big{(}\tilde{U}^{\boldsymbol{k}}_{g}\Big{)}^{-1}=\tilde{H}(p_g\boldsymbol{k}),
\end{align}
where $\tilde{U}^{\boldsymbol{k}}_{g}$ is a symmetry representation of $g$.
Since $\tilde{H}(\bm{k})$ has chiral symmetry $\Gamma = \mathrm{diag}(\bm{1}_N, -\bm{1}_N)$,
we specify a relation between $\tilde{U}^{\boldsymbol{k}}_{g}$ and $\Gamma$ by using a projective factor $c_g$,
which is defined as \cite{PhysRevB.104.035424}
\begin{equation}
    \tilde{U}^{\boldsymbol{k}}_{g}\Gamma=c_g\Gamma \tilde{U}^{\boldsymbol{k}}_{g}.
\end{equation}
Because $\Gamma^2=1$, the projective factor is given by
$c_g\in\{\pm1\}$.
Therefore, symmetry representations commute or anticommute with $\Gamma$.
Then, crystal symmetries can be represented by \cite{PhysRevB.104.035424} 
\begin{align}\label{sup:symmetry_rep}
    \tilde{U}^{\boldsymbol{k}}_{g}=\begin{cases}
        \begin{pmatrix}
        A^{\boldsymbol{k}}_{g}& \\
        & B^{\boldsymbol{k}}_{g}
        \end{pmatrix} & (c_g =1),  \\
        \\
        \begin{pmatrix}
         & B^{\boldsymbol{k}}_{g} \\
        A^{\boldsymbol{k}}_{g}& 
        \end{pmatrix} & (c_g =-1),
    \end{cases}
\end{align}
where $A^{\boldsymbol{k}}_{g}$ and $B^{\boldsymbol{k}}_{g}$ are unitary matrices that are periodic in the Brillouin zone.
By substituting Eq.(\ref{sup:symmetry_rep}) into Eq.(\ref{eq:extended_hermitian_symmetry}), the conditions for the non-Hermitian Hamiltonian can be obtained as follows:
\begin{align}\label{eq:symmetry_non_hermitian_ab}
	A^{\boldsymbol{k}}_{g} H(\boldsymbol{k})
	=\begin{cases}
		H(p_{g}\boldsymbol{k})B^{\boldsymbol{k}}_{g}\ \ (c_g=1), \\
		H^{\dagger}(p_{g}\boldsymbol{k})B^{\boldsymbol{k}}_{g} \ \ (c_g=-1).
	\end{cases}
\end{align}
In this paper, we show that screw symmetry with $c_g=-1$ enforces NHSEs,
and therefore we focus on $c_g=-1$ in the main text.
We discuss the case $c_g=+1$ in Appendix~\ref{appendix:cg=1}.

Next, we apply the above discussion to screw symmetry
to derive the symmetry condition in Eq.~(\ref{eq:generalized_screw_symmetry}),
which corresponds to the screw symmetry with $c_g=-1$.
When a doubled Hermitian Hamiltonian has screw symmetry $g=\{C_{2x}|\hat{\boldsymbol{x}}/2\}$, 
it satisfies 
\begin{align}
\tilde{U}^{\boldsymbol{k}}_{g}\tilde{H}(\boldsymbol{k})\Big{(}\tilde{U}^{\boldsymbol{k}}_{g}\Big{)}^{-1}=\tilde{H}(C_{2x}\boldsymbol{k}), 
\end{align}
where $\tilde{U}^{\boldsymbol{k}}_{g}$ is a representation matrix of the screw symmetry. Because $\{ C_{2x} | \hat{\boldsymbol{x}}/2 \}{^2}=\{ C_{2x}^2 | \hat{\boldsymbol{x}} \}$, it follows that $\tilde{U}^{C_{2x} \boldsymbol{k}}_{g}\tilde{U}^{\boldsymbol{k}}_{g}= e^{-ik_x}(-e^{-ik_x}$) for $C_{2x}^{2}=1(-1)$. 
Therefore, we obtain the following equation:
\begin{gather}
	c_sB^{\boldsymbol{k}}_{g}= \bigl(c_s e^{ik_x}A^{C_{2x}\boldsymbol{k}}_{g}\bigr)^{\dagger}\  ( c_{g}=-1 ), \label{supeq:generalized_screw_B_A_relation_classA}
\end{gather}
where $c_s=1(i)$ for $C_{2x}^{2}=1(-1)$, and  $A^{\boldsymbol{k}}_{g}$ and $B^{\boldsymbol{k}}_{g}$ are unitary matrices in Eq.~(\ref{sup:symmetry_rep}). 
By combining Eq.~(\ref{eq:symmetry_non_hermitian_ab}) with  Eq.~(\ref{supeq:generalized_screw_B_A_relation_classA}), we can obtain
\begin{gather}
	c_s A^{\boldsymbol{k}}_{g}
	H(\boldsymbol{k})=\Bigl(c_s e^{ik_x}A^{C_{2x}\boldsymbol{k}}_{g}
	H(C_{2x}\boldsymbol{k}) \Bigr)^{\dagger}\ \ (c_{g}=-1). \label{eq:non-Hermitian_symmetry}
\end{gather}
By using a unitary matrix  $U^{\boldsymbol{k}}_{g}:=c_s e^{i\frac{k_x}{2}}A^{\boldsymbol{k}}_g$, we can rewrite Eq.~(\ref{eq:non-Hermitian_symmetry}) as Eq.~(\ref{eq:generalized_screw_symmetry}). 

Furthermore, we derive Eq.~(\ref{eq:classAIIdagger_trs_ag}), which indicates the relationship between the unitary matrix $U_{g}^{\boldsymbol{k}}$ and time-reversal symmetry in class class ${\rm AII}^{\dagger}$.
The non-Hermitian Hamiltonian $H(\boldsymbol{k})$ has time-reversal symmetry defined as \cite{PhysRevX.9.041015}
\begin{equation}\label{sup:TRS_aiidagger}
	\mathcal{T}H^{T}(\boldsymbol{k})\mathcal{T}^{-1}=H(-\boldsymbol{k}),
 \end{equation}
 where $\mathcal{T}$ is a unitary matrix and satisfies 
 \begin{equation}
	\mathcal{T} \mathcal{T}^{*}=-1.
\end{equation}
The doubled Hermitian Hamiltonian $\tilde{H}(\boldsymbol{k})$ can obtain time-reversal symmetry given by
\begin{equation}
	\tilde{\mathcal{T}}\tilde{H}(\boldsymbol{k})\tilde{\mathcal{T}}^{-1}=\tilde{H}(-\boldsymbol{k}), 
\end{equation}
where $\tilde{\mathcal{T}}$ is given by
\begin{align}
	\tilde{\mathcal{T}}=\begin{pmatrix}
		& \mathcal{T} \\
		\mathcal{T} &
	\end{pmatrix}K,
\end{align}
and satisfies $\tilde{\mathcal{T}}^{2}=-1$. 
The symmetry class of $\tilde{H}(\boldsymbol{k})$ is class DIII \cite{PhysRevB.55.1142}
because $\tilde{H}(\bm{k})$ has particle-hole symmetry that satisfies $\tilde{\mathcal{P}}^2=1$ owing to $\Gamma \tilde{\mathcal{T}} = -\tilde{\mathcal{T}}\Gamma $.  
We also obtain $\tilde{\mathcal{T}} \tilde{U}^{\boldsymbol{k}}_{g}  = \tilde{U}^{-\boldsymbol{k}}_{g}   \tilde{\mathcal{T}}$
by imposing the commutation relation on $\{ C_{2x} | \hat{\boldsymbol{x}}/2 \}$ and $\tilde{\mathcal{T}}$.
Therefore, we have 
\begin{align}
	\mathcal{T} 
	(A^{\boldsymbol{k}}_{g})^{*}
	=B^{-\boldsymbol{k}}_{g} \mathcal{T}.
\end{align} 
Moreover, we choose $c_s=i$ because $\tilde{\mathcal{T}}^{2}=-1$.
Hence, we obtain
\begin{align}\label{sup:eq_ta_eat}
	\mathcal{T} 
	(A^{\boldsymbol{k}}_{g})^{*}
	=- e^{ik_x}
	(A^{-C_{2x}\boldsymbol{k}}_{g})^{-1} \mathcal{T}.
\end{align}
By using the unitary matrix  $U_{g}^{\boldsymbol{k}}=i e^{i\frac{k_x}{2}}A^{\boldsymbol{k}}_{g}$, we get Eq.~(\ref{eq:classAIIdagger_trs_ag}).

\subsection{Mechanism of NHSEs enforced by nonsymmorphic symmetry in class AII$^{\dagger}$}

As shown in Sec.~\ref{sec:condition_for_NHSE}, 
 if $H(\boldsymbol{k})$ satisfies the condition in Eq.~(\ref{eq:generalized_screw_symmetry}), 
$\tilde{H}(\boldsymbol{k})$ has the screw rotation symmetry $g=\{ C_{2x} | \hat{\boldsymbol{x}}/2 \}$. 
Thus, eigenstates of $\tilde{H}(\boldsymbol{k})$ have the screw rotation eigenvalues $\pm c_s e^{-i\frac{k_x}{2}}$ on the screw-invariant lines ($k_{y}=0, \pi$). 
In order to take time-reversal symmetry into account,
we consider a case where Eq.~(\ref{eq:classAIIdagger_trs_ag}) holds with $C_{2x}^2=-1$ ($c_s=i$).

We show that the doubled Hermitian Hamiltonian has 2D Weyl points at zero energy if $c_g=-1$. The eigenstates on the screw-invariant lines ($k_y =0,\pi$) are labeled by the screw symmetry eigenvalues $\pm i e^{-i\frac{k_x}{2}}$. Because of the time-reversal symmetry,
while Kramers pairs  
at $k_{x}=0$ have opposite screw rotation eigenvalues $\pm i$, 
Kramers pairs at $k_{x}=\pi$ have the same eigenvalues [Fig.~\ref{fig:concept_AIIdagger}(a)].
Moreover, because the chiral operator anticommutes with the screw operator, 
a pair of eigenstates with the energy $\mathcal{E}$ and $-\mathcal{E}$ has screw eigenvalues with the opposite signs \cite{comment2}.
Since the eigenvalues must cross on the line,
the eigenstates with these symmetry eigenvalues enforce Weyl points on the screw-invariant line.  
Because the number of bands of the doubled Hermitian Hamiltonian is $8M+4$ by assumption, the chiral symmetry fixes the 2D Weyl points at zero energy.

The emergent Weyl points give a path with a nonzero winding number in the Brillouin zone,
depending on their chirality and positions \cite{Matsuura_2013, RevModPhys.88.035005,  PhysRevLett.118.156401, PhysRevB.105.035429, PhysRevB.106.045126}. In addition, the winding number satisfies $W_x(-k_y)=-W_x(k_y)$ because $\mathcal{T}H^{T}(\boldsymbol{k})\mathcal{T}^{-1}=H(-\boldsymbol{k})$.  
Also, as we show in Appendix~\ref{appendix:review_doubled_winding}, when the time-reversal and the chiral operators anticommute, the 2D Weyl points at $\boldsymbol{k}_{0}$ and $-\boldsymbol{k}_0$ have the same chirality [see Eq.~(\ref{eq:trs_winding_conclusion}) in Appendix~\ref{appendix:review_doubled_winding}].
Thus, a time-reversal pair of the Weyl points on the screw-invariant line has the same chirality.
It follows that the winding number $W_x (k_y)$ is nonzero at any $k_y$ due to the positions and the chirality of the Weyl points [Fig.~\ref{fig:concept_AIIdagger}(b)].
 Furthermore, we can find anisotropy in the winding numbers $W_x(k_y)$ and $W_y(k_x)$. Because we cannot annihilate the Weyl points on the screw-invariant lines, we can always find a path for a nonzero $W_x(k_y)$ in the Brillouin zone. On the other hand,  $W_y(k_x)$ can be zero in the presence of the Weyl points with the opposite chirality at the same $k_x$.
 As a result, the NHSE under the OBC in the $x$ direction is more robust than that under the OBC in the $y$ direction.

	Moreover, since the 2D Weyl points appear at zero energy,
	we obtain ${\rm det}[ H(\boldsymbol{k})]{\rm det}[ H^{\dagger}(\boldsymbol{k})]=0$ at some wavevector on the screw-invariant lines.
	Therefore, the point-gap closing of $H(\boldsymbol{k})$ is found at $E=0$ under the full PBC.
	Thus, the nonsymmorphic-symmetry-enforced NHSE always occurs with the point-gap closing on the screw-invariant lines.

 	\begin{table}
		\begin{tabular}{c c c}
			\hline
			\hline
			\begin{tabular}{c}Symmetry class\end{tabular} &\begin{tabular}{c} Matrix size of \\  $H(\boldsymbol{k})$ and $A_g^{\boldsymbol{k}}$\end{tabular}  & \begin{tabular}{c} Relation between \\ $\mathcal{T}$ and  $U_{g}^{\boldsymbol{k}}$  \end{tabular} \\ 
			\hline
			class A & $2M+1$ 
			& -- 
			\\ 
			class AII  & $4M+2$ & 
			$ \mathcal{T} (U_{g}^{\boldsymbol{k}})^*= -U_{g}^{-\boldsymbol{k}}  \mathcal{T}
			$
			\\
			class AII$^{\dagger}$  & $4M+2$  & 
            $\mathcal{T}(U_g^{\boldsymbol{k}})^{*}=-(U_g^{-C_{2x}\boldsymbol{k}})^{\dagger}\mathcal{T}$
			\\
			\hline
			\hline
		\end{tabular}
		\caption{Conditions of the matrix size and the symmetries for nonsymmorphic-symmetry-enforced NHSEs. When a given non-Hermitian Hamiltonian satisfies the conditions of the symmetry and the matrix size, the system can realize symmetry-enforced NHSEs.
			}	\label{tab:summary_2d_nonsymmorphic_main}
	\end{table}
 
	\subsection{Nonsymmorphic-symmetry-enforced NHSE in other symmetry classes}
	Next, we discuss nonsymmorphic-symmetry-enforced NHSEs in symmetry class A and class AII.
	Non-Hermitian  Hamiltonians in class A do not have any internal symmetries.
    Meanwhile, non-Hermitian Hamiltonians $H(\boldsymbol{k})$ in symmetry class AII have time-reversal symmetry 
	$
	\mathcal{T}H^{*}(\boldsymbol{k})\mathcal{T}^{-1}=H(-\boldsymbol{k}),
	$
	where $\mathcal{T}$ is a unitary matrix satisfying $\mathcal{T}\mathcal{T}^{*}=-1$ \cite{PhysRevX.9.041015}. 
    While Eq.~(\ref{eq:generalized_screw_symmetry}) is also applicable to class A and class AII,
    $U_{g}^{\boldsymbol{k}}$ in class AII needs to satisfy a relationship with $\mathcal{T}$ in Table \ref{tab:summary_2d_nonsymmorphic_main}.

		Furthermore, we discuss the matrix size of non-Hermitian Hamiltonians $H(\boldsymbol{k})$ to realize symmetry-enforced NHSEs.
	To see the conditions of the matrix size
	we investigate Weyl points of the doubled Hermitian Hamiltonians $\tilde{H}(\boldsymbol{k})$ with the screw symmetry and with/without time-reversal symmetry.
	The details of the band topology are given in \textcolor{black}{Appendices \ref{sm:symmetry_condition} and \ref{appendx:classAII}.}
 
	When the matrix size of $H(\boldsymbol{k})$ in class A is $2M+1$, $\tilde{H}(\boldsymbol{k})$ has a 2D Weyl point on the screw invariant lines. On the other hand, when the matrix size of $H(\boldsymbol{k})$ in class AII is $4M+2$, $\tilde{H}(\boldsymbol{k})$ has a pair of 2D Weyl points on the screw invariant lines. 
    Therefore,
    nonsymmorphic-symmetry-enforced NHSEs can happen in the two classes (see Appendices~\ref{sm:symmetry_condition} and \ref{appendx:classAII}).
	These results are summarized in Table \ref{tab:summary_2d_nonsymmorphic_main}.
	
   In contrast, NHSEs are not enforced in 2D class AI and 2D class AI$^\dagger$.
    We show the absence of NHSEs enforced by screw symmetry in the two classes.
    The detail is given in Appendix~\ref{appendx:classAI_and_AIdagger}. 

	\begin{figure}
		\includegraphics[width=1.\columnwidth]{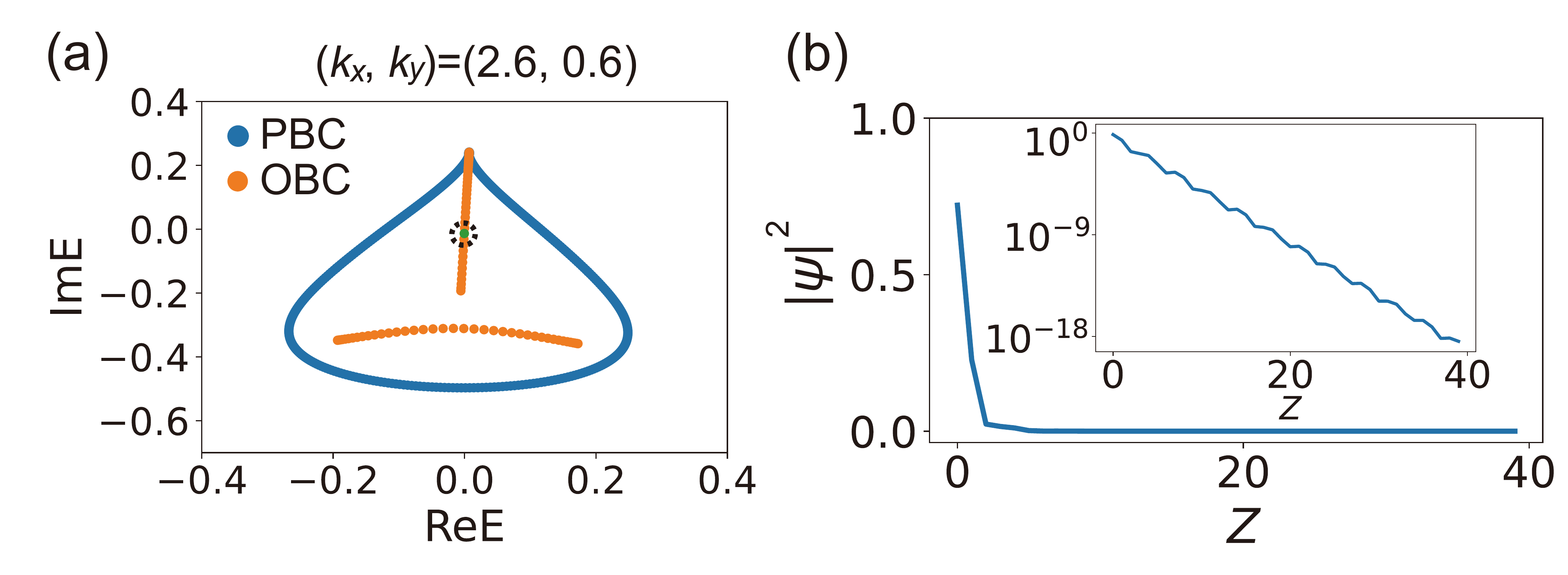}
		\caption{Nonsymmorphic-symmetry-enforced NHSEs in the 3D system. We choose the parameters $t_{0}=0.125$, $t_a=0.1$,  $t_{b}=t_{d}=0.05$, and $t_{c}=0.025$. 
			(a) The energy spectra of $H_{\rm 3D}(\boldsymbol{k})$ with $(k_x, k_y)=(2.6,0.6)$ under the PBC in the $z$ direction and the OBC in the $z$ direction. 
            The system size in the $z$ direction is 40. (b) The spatial distributions of the skin mode in the dotted circle in (a). The inset shows the spatial distribution in log scale.} 
		\label{fig:skin_eff}
	\end{figure}
	
\section{3D NHSE enforced by nonsymmorphic symmetry}
While a nonsymmorphic-symmetry-enforced NHSE cannot be realized in 2D systems in class AI,
it can occur in 3D class AI.
To show this, let us extend the above discussion for 2D systems to that for 3D systems.
We introduce a 3D two-band non-Hermitian model on a primitive orthorhombic lattice.
	The model is given by 
	\begin{equation}
		H_{\rm 3D}(\boldsymbol{k})= v_{0}(\boldsymbol{k}) \sigma_0+ \sum_{i=x,y} v_i (\boldsymbol{k}) \sigma_i,
	\end{equation}
	where $\sigma _i$ ($i=x,y,z$) and $\sigma_0$ are the Pauli matrices and the identity matrix, respectively. Here, the coefficient functions are given by 
	\begin{align}
		v_{0}(\boldsymbol{k})=&t_{0} f_{1}(\boldsymbol{k}_{\parallel})-it_{a} \sin k_{z} f_{1}(\boldsymbol{k}_{\parallel}), \nonumber \\
		v_{x}(\boldsymbol{k})=& -it_{b}\sin k_{z} f_{1}(\boldsymbol{k}_{\parallel})-it_{c}\sin k_{z}f_{2}(\boldsymbol{k}_{\parallel}) \nonumber \\
		&+t_{d}(1-\cos k_{z})f_{2}(\boldsymbol{k}_{\parallel}), \nonumber \\
		v_{y}(\boldsymbol{k})=&-it_{b} (1-\cos k_{z})f_{1}(\boldsymbol{k}_{\parallel})+it_{c}(1+\cos k_{z}) f_{2}(\boldsymbol{k}_{\parallel}) \nonumber \\
		&-t_{d} \sin k_z f_2 (\boldsymbol{k}_{\parallel}), 
	\end{align}
	where $\boldsymbol{k}_{\parallel}= (k_{x}, k_{y})$ is the $x$-$y$ plane momentum,  $f_{1}(\boldsymbol{k}_{\parallel}) = (1+e^{-ik_{x}})(1+e^{-ik_{y}})$, and $f_{2}(\boldsymbol{k}_{\parallel}) = (e^{ik_{x}}-e^{-2ik_{x}})(1-e^{-ik_{y}})-(e^{ik_{y}}-e^{-2ik_{y}})(1-e^{-ik_{x}})$. 
	This model has time-reversal symmetry:  $H_{\rm 3D}^{*}(\boldsymbol{k})=H_{\rm 3D}(-\boldsymbol{k})$, and therefore it belongs to class AI. 
	We calculate energy spectra of $H_{\rm 3D}(\boldsymbol{k})$ under the PBC in both the $x$ and the $y$ directions.
	Figure~\ref{fig:skin_eff}(a) shows that the NHSE occurs under the OBC in the $z$ direction.
	Then Fig.~\ref{fig:skin_eff}(b) shows that the eigenstates are localized at the surface normal to the $z$ direction.

To see that the NHSE stems from nonsymmorphic symmetry,
we study the doubled Hermitian Hamiltonian $\tilde{H}_{\rm 3D}(\boldsymbol{k})$ obtained from $H_{\rm 3D}(\boldsymbol{k})$ by using Eq.~(\ref{eq:extended_hermitian_2d}).
The doubled Hermitian Hamiltonian $\tilde{H}_{\rm 3D}(\boldsymbol{k})$ is given by 
\begin{widetext}
\begin{align}
    \tilde{H}_{\rm 3D}(\boldsymbol{k})=& V_{1}(\boldsymbol{k}_{\parallel}) \biggl(
    \frac{t_0}{4}\sin{k_z}\tau_{y}\otimes \sigma_0 + \frac{t_a}{4} \tau_x \otimes \sigma_0 +\frac{t_b}{8} \sin k_z \tau_y \otimes \sigma_x + \frac{t_b}{8}(1-\cos k_x) \tau_y \otimes \sigma_y \biggr) \nonumber \\
    +& V_{2}(\boldsymbol{k}_{\parallel}) \Bigl( -\frac{t_0}{4}\sin{k_z}\tau_{x}\otimes \sigma_0 + \frac{t_a}{4} \tau_y \otimes \sigma_0  -\frac{t_b}{8} \sin k_z \tau_x \otimes \sigma_x -\frac{t_b}{8}(1-\cos k_z)\tau_x \otimes \sigma_y
    \biggr)\nonumber \\
    +& V_{3}(\boldsymbol{k}_{\parallel})\Bigl(
    -\frac{t_c}{8}(1+\cos k_z)\tau_y \otimes \sigma_y +\frac{t_c}{8}\sin k_z \tau_y \otimes \sigma_x -\frac{t_d}{8}\sin k_z \tau_x \otimes \sigma_y +\frac{t_d}{8}(1-\cos k_z)\tau_x \otimes \sigma_x
    \Bigr)\nonumber \\
     +&V_4(\boldsymbol{k}_{\parallel}) \Bigl(
    -\frac{t_c}{8}(1+\cos k_z)\tau_x \otimes \sigma_y +\frac{t_c}{8}\sin k_z \tau_x \otimes \sigma_x +\frac{t_d}{8}\sin k_z \tau_y \otimes \sigma_y -\frac{t_d}{8}(1-\cos k_z)\tau_y \otimes \sigma_x \Bigr), 
\end{align}
with 
\begin{align}
V_{1}(\boldsymbol{k}_{\parallel})=& 1+\cos{k_x}+\cos{k_y}+\cos (k_x + k_y), \nonumber \\
V_{2}(\boldsymbol{k}_{\parallel})=&\sin{k_x}+\sin{k_y}+\sin (k_x + k_y), \nonumber \\
V_{3}(\boldsymbol{k}_{\parallel})=& (\cos k_x -\cos 2k_x )(1-\cos k_y) -(\cos k_y -\cos 2k_y ) (1-\cos k_x)-(\sin k_x + \sin 2k_x )\sin k_y  \nonumber \\
    &+(\sin k_y + \sin 2k_y )\sin k_x,  \nonumber  \\ 
    V_{4}(\boldsymbol{k}_{\parallel})= &(\cos k_x -\cos 2k_x )\sin k_y +(\sin k_x + \sin 2k_x )(1-\cos k_y)-(\cos k_y -\cos  2k_y )\sin k_x  \nonumber \\
    & -(\sin k_y + \sin 2k_y )(1-\cos k_x),
\end{align}
\end{widetext}
where $\tau_i$ ($i=x,y,z$) and $\tau _0$ are the Pauli matrices and the identity matrix. 
The doubled Hermitian Hamiltonian corresponds to a 3D tight-binding model on a primitive orthorhombic lattice with the four sublattices at $\boldsymbol{r}_{\alpha}=(1/4,1/4,0)$, $(1/4,1/4,1/2)$,  $(3/4,3/4,0)$, and $(3/4,3/4,1/2)$ in each unit cell \cite{PhysRevB.96.155206}.
In the model, the Pauli matrices $\sigma_i$ represent $z=0$ and $z=1/2$ sublattices, and the Pauli matrices $\tau_i$ represent the $(x,y)=(1/4, 1/4)$ and $(x,y)=(3/4, 3/4)$ sublattices. 	

\begin{table*}
\begin{tabular}{c c c c c}
\hline
\hline
Symmetry operation &$c_g$  & $A_{g}^{\boldsymbol{k}}$ & $B_{g}^{\boldsymbol{k}}$ &Symmetry relation \\ 
\hline
$\{ I|0 \}$ &
$-1$\ \ \ 
& $e^{-i(k_x+k_y+k_z)}\sigma_x$\ \ \  & $e^{-i(k_x+k_y+k_z)}\sigma_x$\ \ \   & 
$A_{g}^{\boldsymbol{k}}H_{\rm 3D}(k_x, k_y, k_z)=H^{\dagger}_{\rm 3D}(-k_x, -k_y, -k_z)B_{g}^{\boldsymbol{k}}$
 \\ 
$\{C_{4z}|\tfrac{1}{2} \hat{\boldsymbol{x}} \}$ & 1 & $\sigma_0$ & $e^{-ik_y}\sigma_0$ & $A_{g}^{\boldsymbol{k}}H_{\rm 3D}(k_x,k_y,k_z)=H_{\rm 3D}(-k_y,k_x,k_z) B_{g}^{\boldsymbol{k}}$\\ 
$\{C_{2y}|\tfrac{1}{2} \hat{\boldsymbol{y}}+\tfrac{1}{2} \hat{\boldsymbol{z}} \}$ & $-1$ &
$ \begin{pmatrix}
e^{-ik_x} & \\
& e^{-i(k_x+k_z)}
\end{pmatrix}$ 
& 
$ \begin{pmatrix}
e^{-i(k_x+k_y)} & \\
& e^{-i(k_x+k_y+k_z)}
\end{pmatrix}$ 
& $A_{g}^{\boldsymbol{k}}H_{\rm 3D}(k_x,k_y,k_z)=H^{\dagger}_{\rm 3D}(-k_x,k_y,-k_z) B_{g}^{\boldsymbol{k}}$\\
$\{C_{2z}|\tfrac{1}{2} \hat{\boldsymbol{x}}+\tfrac{1}{2} \hat{\boldsymbol{y}} \}$ & 1 & $\sigma_0$ & $e^{-i(k_x+k_y)}\sigma_0$ & $A_{g}^{\boldsymbol{k}}H_{\rm 3D}(k_x,k_y,k_z)=H_{\rm 3D}(-k_x,-k_y,k_z)B_{g}^{\boldsymbol{k}}$\\ 
$\tilde{\mathcal{T}}$ & -- & -- & -- & $H^{*}_{\rm 3D}(k_x, k_y, k_z)=H_{\rm 3D}(-k_x, -k_y, -k_z)$
\\
\hline
\hline
\end{tabular}
\caption{The symmetry relations of the 3D non-Hermitian Hamiltonian ${H}_{\rm 3D}(\boldsymbol{k})$. }
\label{tab:3dsymmetry}
\end{table*}	

To see the band topology, 
we discuss symmetries of the doubled Hermitian Hamiltonian $\tilde{H}_{\rm 3D}(\boldsymbol{k})$ 
by using the Seitz notation.
Let $\bm{x}$, $\bm{y}$, and $\bm{z}$ denote the lattice vectors in the $x$, $y$, and $z$ directions, respectively.
The doubled Hermitian Hamiltonian belongs to space group $P4/ncc$  (No.~130) generated by $\{ I| \boldsymbol{0} \}$,  $\{C_{4z}|\tfrac{1}{2} \hat{\boldsymbol{x}} \}$, $ \{C_{2y}|\tfrac{1}{2} \hat{\boldsymbol{y}}+\tfrac{1}{2} \hat{\boldsymbol{z}} \}$, and $\{C_{2z}|\tfrac{1}{2} \hat{\boldsymbol{x}}+\tfrac{1}{2} \hat{\boldsymbol{y}} \}$, 
where $I$ represents the inversion, and $C_{4z}$ represents the fourfold rotation around the $z$ axis, and $C_{2y}$ and $C_{2z}$ represents the twofold rotation around the $y$ and the $z$ axes, respectively.
The symmetry representations $\tilde{U}^{\boldsymbol{k}}_{g}$ are
\begin{align}
&\tilde{U}^{\boldsymbol{k}}_{\{ I| \boldsymbol{0} \}}= e^{-i(k_{x}+k_{y}+k_{z})}\tau_{x} \otimes \sigma_{x}, \nonumber \\
&\tilde{U}^{\boldsymbol{k}}_{\{C_{4z}|\tfrac{1}{2} \hat{\boldsymbol{x}} \}}=\begin{pmatrix}
1&  \\
&e^{-ik_{y}}
\end{pmatrix}\otimes \sigma_{0}, \nonumber \\
&\tilde{U}^{\boldsymbol{k}}_{\{C_{2y}|\tfrac{1}{2} \hat{\boldsymbol{y}}+\tfrac{1}{2} \hat{\boldsymbol{z}} \} }=\begin{pmatrix}
& e^{-i(k_{x}+k_{y})}  \\
e^{-ik_{x}} &
\end{pmatrix}\otimes
\begin{pmatrix}
1 & \\
& e^{-ik_{z}}
\end{pmatrix}
, \nonumber \\
&\tilde{U}^{\boldsymbol{k}}_{\{C_{2z}|\tfrac{1}{2} \hat{\boldsymbol{x}}+\tfrac{1}{2} \hat{\boldsymbol{y}} \}}=\begin{pmatrix}
1&   \\
 & e^{-i(k_{x}+k_{y})}
\end{pmatrix}\otimes \sigma_{0}.
\end{align}
We can understand that the crystal symmetries stem from symmetry relations in Eq.~(\ref{sup:symmetry_rep}).
Their projective factors can be calculated from the symmetry representations.
The relations are summarized in  Table~\ref{tab:3dsymmetry}.
Furthermore, the model $H_{\rm 3D}(\boldsymbol{k})$ has time-reversal symmetry 
$H_{\rm 3D}^*(\boldsymbol{k})=H_{\rm 3D}(-\boldsymbol{k})$,
which means that it belongs to class AI.
Thus, $\tilde{H}_{\rm 3D}(\boldsymbol{k})$ has time-reversal symmetry $\tilde{\mathcal{T}}\tilde{H}_{\rm 3D}(\boldsymbol{k})\tilde{\mathcal{T}}^{-1}=\tilde{H}_{\rm 3D}(-\boldsymbol{k})$
with $\tilde{\mathcal{T}}=K$, which satisfies $\tilde{\mathcal{T}}^2=1$.

\begin{figure}[b]
\includegraphics[width=1.\columnwidth]{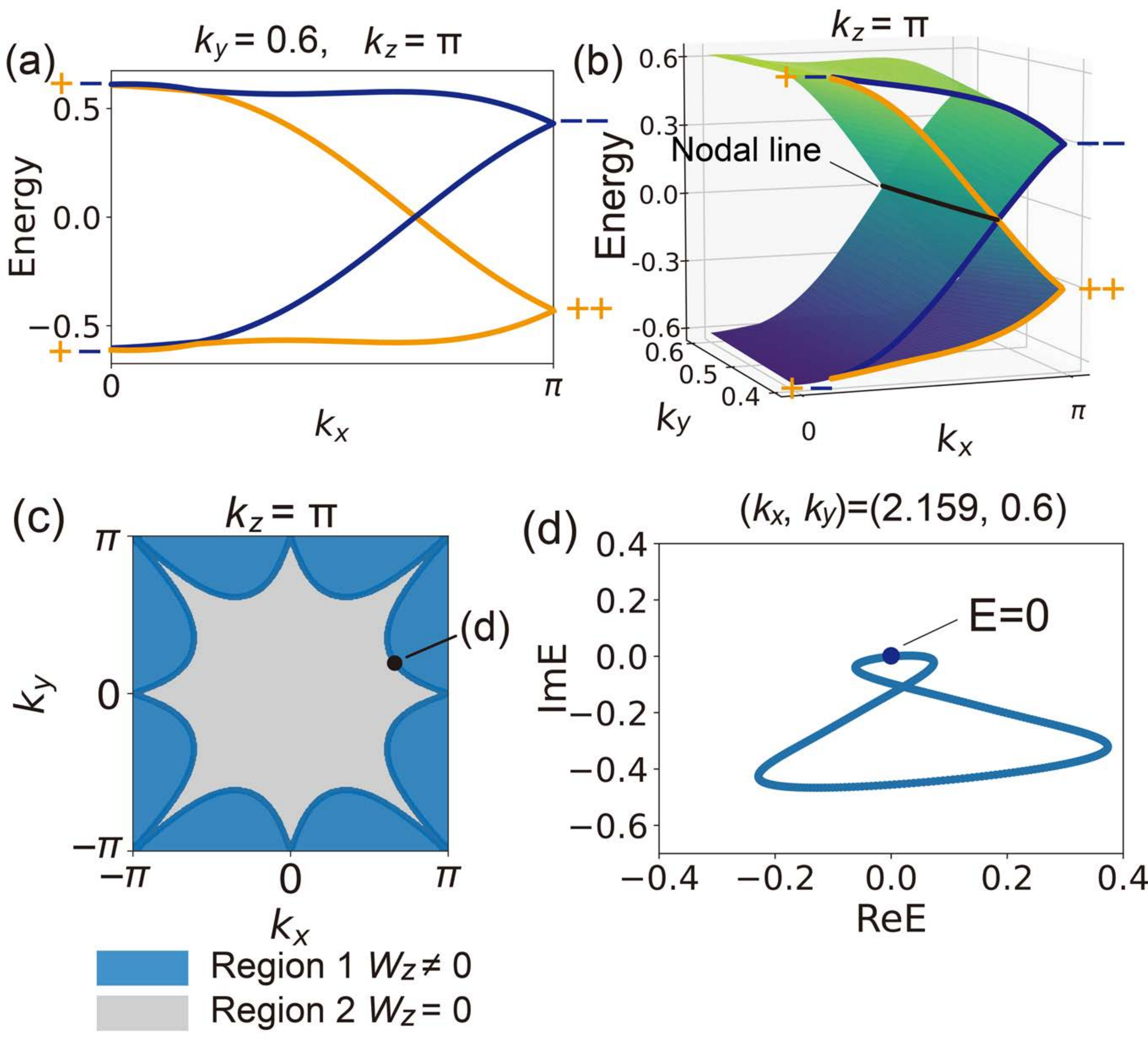}
\caption{
(a) The energy spectra of the doubled Hermitian Hamiltonian $\tilde{H}_{\rm 3D}(\boldsymbol{k})$ on the $(k_y, k_z)=(0.6, \pi)$ line. $\pm$ denotes the sign of the glide symmetry eigenvalues $\pm e^{-\frac{i}{2}(k_x+k_y)}$. (b) The band structures of the doubled Hermitian Hamiltonian $\tilde{H}_{\rm 3D}(\boldsymbol{k})$ on the $k_z=\pi$ plane.
(c) Top view of the winding number $W_{z}(\boldsymbol{k}_{\parallel})$ on the $k_{x}$-$k_{y}$ space. The winding number varies on the nodal line on the $k_z=\pi$ plane. (d) The energy spectra of the non-Hermitian Hamiltonian  ${H}_{\rm 3D}(\boldsymbol{k})$ with $(k_x, k_y)=(2.159,0.6)$ calculated along the $k_z$ direction under the full PBC. The parameters are the same as those in Fig.~\ref{fig:skin_eff}
} 
\label{fig:hourglass_band}
\end{figure}

Importantly, when time-reversal symmetry with $\tilde{\mathcal{T}}^2=1$ is present,
the space group $P4/ncc$ realizes nodal lines enforced by glide reflection symmetry \cite{PhysRevB.96.155206}.
The glide symmetry is given by
$G=\{ M_z |\frac{1}{2}\hat{\boldsymbol{x}}+\frac{1}{2}\hat{\boldsymbol{y}} \} = \{C_{2z}|\tfrac{1}{2} \hat{\boldsymbol{x}}+\tfrac{1}{2} \hat{\boldsymbol{y}} \}
\{ I| \boldsymbol{0} \}$,
where $M_{z}$ is mirror reflection with the $x$-$y$ mirror plane.
The projective factor of the glide symmetry is $c_G=-1$.
We show that $\tilde{H}_{\rm 3D}(\boldsymbol{k})$ can host a symmetry-enforced nodal line on the glide-invariant plane $k_z=\pi$. 
The eigenstates are labeled by the glide reflection eigenvalues $\pm e^{-\frac{i}{2}(k_x+k_y)}$ on the glide-invariant plane $k_z=\pi$. 
Here, eigenstates at $k_x=0$ and $\pi$ are necessarily doubly degenerate in the space group because of combinations of time-reversal and crystal symmetries \cite{PhysRevB.96.155206}. 
The doubly-degenerate states at $k_x=0$ ($\pi$)  have the opposite (same)  glide eigenvalues. 
Since $c_G=-1$, a pair of eigenstates with the energies $\mathcal{E}$ and $-\mathcal{E}$ have glide eigenvalues of the opposite sign [Fig.~\ref{fig:hourglass_band}(a)].
Figure~\ref{fig:hourglass_band}(a) shows that the glide eigenvalues need to cross between $k_x=0$ and $k_x=\pi$ on the glide-invariant  $k_x$-$k_y$ plane. 
The eigenvalue crossing enforces band degeneracy at zero energy due to chiral symmetry because the matrix size of $\tilde{H}_{\rm 3D}(\boldsymbol{k})$ is $4$. 
In general, the symmetry-enforced nodal line can appear at zero energy if the matrix size is $8M+4$.
The mechanism is similar to symmetry-enforced 2D Weyl point in time-reversal-invariant classes with $\mathcal{T}^2=-1$ in Fig.~\ref{fig:concept_AIIdagger}.
The degeneracy forms the nodal line on the glide-invariant plane $k_z=\pi$ [Fig.~\ref{fig:hourglass_band}(b)].
	
Nodal lines generally lead to a path with a nontrivial winding number in the 3D Brillouin zone \cite{PhysRevB.84.235126, PhysRevB.90.115111, PhysRevB.92.081201,doi:10.1063/1.4926545, PhysRevLett.115.036806}.
	In this case, a winding number calculated along the $k_z$ direction changes its value on the nodal line.
	Thus, the corresponding non-Hermitian Hamiltonian $H_{\rm 3D}(\boldsymbol{k})$ has a nonzero winding number $W_{z}(\boldsymbol{k}_{\parallel})$, as shown in Fig.~\ref{fig:hourglass_band}(c).
	As a result, we see that a symmetry-enforced NHSE happens in the 3D non-Hermitian lattice. Also, we find that the point gap of the non-Hermitian Hamiltonian $H_{\rm 3D}(\boldsymbol{k})$ closes at  $E=0$ on
	specific lines along the $k_z$ direction under the full PBC [Fig.~\ref{fig:hourglass_band}(d)]. 
	We can also grasp the point-gap closing at $E=0$ from the symmetry-enforced nodal line.
	
	\section{Conclusion and discussion}
	In summary, we study NHSEs enforced by nonsymmorphic symmetry in 2D and 3D non-Hermitian systems, which we call nonsymmorphic-symmetry-enforced NHSEs.
	We have clarified the conditions of symmetry and matrix size of non-Hermitian Hamiltonians for the NHSEs.
	If the conditions are satisfied, these  NHSEs inevitably happen. 
    We have also revealed that the point gap at $E=0$ closes under the full PBC when the NHSEs occur. 
    These NHSEs can occur in 2D systems not only  without time-reversal symmetry but also with time-reversal symmetry.

	NHSEs have been experimentally observed in artificially designed systems \cite{xiao2020non, weidemann2020topological, helbig2020generalized, brandenbourger2019non, chen2021realization, zhang2021acoustic, PhysRevResearch.2.023265, palacios2021guided, shang2022experimental}. We believe that our theory can be readily confirmed in artificial two-dimensional systems, such as active matter  \cite{palacios2021guided} and electric circuits \cite{PhysRevResearch.2.023265, shang2022experimental}.  
    Symmetries in these artificial platforms can be flexibly controlled.
    Our theory is helpful for design of non-Hermitian band structures in terms of spatial symmetries.

	\begin{acknowledgments}
		This work was supported by Japan Society for the Promotion of Science (JSPS) KAKENHI Grants No.~JP20H01845, No.~JP21J22264, and No.~JP23K13033. 
	\end{acknowledgments}

\appendix

\section{Review of doubled Hermitian Hamiltonians and winding numbers}\label{appendix:review_doubled_winding}
\subsection{Winding numbers for non-Hermitian Hamiltonians and doubled Hermitian Hamiltonians}
We briefly review the relationship between the NHSEs and band topology of a doubled Hermitian Hamiltonian.
Throughout this work, we choose the convention for non-Hermitian Hamiltonians $H(\bm{k})$ to satisfy $H(\bm{k}+\bm{G})=H(\bm{k})$,
as used for the Fourier transformation in the main text.

We start with a winding number for the NHSE in 2D and 3D non-Hermitian systems,
which is defined as \cite{zhang2022universal}
\begin{align}
    W(E)=\frac{1}{2\pi i}\oint_{l}d\bm{k} \cdot \partial_{\boldsymbol{k}} \log \det [H(\boldsymbol{k})-E],
\end{align}
where $l$ is a loop in the Brillouin zone, and $E$ is a reference energy.
If a winding number with a reference energy is nonzero in the system,
a NHSE happens under an OBC \cite{zhang2022universal}.
On the other hand, it is unclear what OBC leads to the NHSE among the various OBCs. We can determine how to take an OBC to observe a NHSE by choosing a line in the Brillouin zone as $l$.
For instance, let us introduce the following winding number 
\begin{align}
    W_i(k_{\perp}, E)=\frac{1}{2\pi i}\int_{0}^{2\pi}dk_i \frac{\partial}{\partial k_i}\log \det [H(\bm{k})-E],
\end{align}
where $k_{\perp}$ is the wave vector perpendiclular to $k_i$ $(i=x,y,z)$.
If $W(k_{\perp }, E)$ is nonzero, a NHSE happens under the OBC in the $i$ direction and under the PBC in the other directions \cite{PhysRevLett.124.086801}.
In this paper, we investigate the winding number with $E=0$.

Next, we see the band topology of a doubled Hermitian Hamiltonian defined by Eq.~(\ref{eq:extended_hermitian_2d}).
The doubled Hermitian Hamiltonian has chiral symmetry
$\Gamma={\rm diag}(\boldsymbol{1}_{N}, -\boldsymbol{1}_{N})$,
where $N$ is the matrix size of ${H}(\boldsymbol{k})$.
For the topological classification of the doubled Hermitian Hamiltonians, we introduce the $Q$ matrix defined as \cite{RevModPhys.88.035005} 
\begin{equation}  
Q(\boldsymbol{k})=1-2\sum_{\mathcal{E}_{\alpha}<0}\ket{\Psi^{\alpha}(\boldsymbol{k})}\bra{\Psi^{\alpha}(\boldsymbol{k})},
\end{equation}
where $\ket{\Psi^{\alpha}(\boldsymbol{k})}$ is an eigenstate of the doubled Hermitian Hamiltonian with the energy $\mathcal{E}_{\alpha}$.
When chiral symmetry is diagonal,
the $Q$ matrix is block off-diagonal, represented by
\begin{align}\label{eq:qmatrix_off_diagonal}
    Q(\boldsymbol{k})=\begin{pmatrix}
    0 & q (\boldsymbol{k}) \\
    q^{\dagger}( \boldsymbol{k}) & 0
    \end{pmatrix},
\end{align}
where $q(\bm{k})$ is a unitary matrix.
When only chiral symmetry is present, the
Hamiltonians belong to class AIII \cite{PhysRevB.55.1142}.
The Hermitian Hamiltonians can be characterized by
\begin{align}\label{smeq:line_winding}
    \nu_{{i}}(k_{\perp})=\frac{1}{2\pi i}\int^{2\pi}_{0} dk_i \frac{\partial}{\partial k_i}  \log \det [q(\boldsymbol{k})],
\end{align}
which is also a winding number \cite{RevModPhys.88.035005}. 
In high-dimensional systems, the winding number $\nu_{{i}}(k_{\perp})$ is used to characterize the topological gapless band structures \cite{RevModPhys.88.035005, PhysRevB.106.045126}.

Hereafter we show that the winding number $W_{{i}}(k_{\perp})$ in the main text is equivalent to the winding number $\nu (k_{\perp})$, according to Ref.~\cite{PhysRevB.106.045126}.
Any doubled Hermitian Hamiltonian $\tilde{H}(\boldsymbol{k})$ commutes with the $Q$ matrix because
\begin{align}
    \tilde{H}(\boldsymbol{k})Q(\boldsymbol{k})
    =&\tilde{H}(\boldsymbol{k})-2\sum_{\mathcal{E}_{\alpha}<0}\mathcal{E}_{\alpha}\ket{\Psi^{\alpha}(\boldsymbol{k})}\bra{\Psi^{\alpha}(\boldsymbol{k})} \nonumber \\
    =&Q(\boldsymbol{k})\tilde{H}(\boldsymbol{k}).
\end{align}
Because the commutation relation can be written on the basis of Eq.~(\ref{eq:qmatrix_off_diagonal}) as 
\begin{align}
&\begin{pmatrix}
    0 & H (\boldsymbol{k}) \\
    H^{\dagger}( \boldsymbol{k}) & 0
    \end{pmatrix} 
    \begin{pmatrix}
    0 & q (\boldsymbol{k}) \\
    q^{\dagger}( \boldsymbol{k}) & 0
    \end{pmatrix}\nonumber \\
    -&
    \begin{pmatrix}
    0 & q (\boldsymbol{k}) \\
    q^{\dagger}( \boldsymbol{k}) & 0
    \end{pmatrix} 
    \begin{pmatrix}
    0 & H (\boldsymbol{k}) \\
    H^{\dagger}( \boldsymbol{k}) & 0
    \end{pmatrix} 
    =0, 
\end{align}
we obtain
\begin{equation}
    H(\boldsymbol{k})q^{\dagger}(\boldsymbol{k})=q(\boldsymbol{k})H^{\dagger}(\boldsymbol{k}).
\end{equation}
Therefore, we have
\begin{align}
    \det [H(\boldsymbol{k})q^{\dagger}(\boldsymbol{k})]&=\det [q(\boldsymbol{k})H^{\dagger}(\boldsymbol{k})]\nonumber \\
    &=\det [H(\boldsymbol{k})q^{\dagger}(\boldsymbol{k})]^{*},
\end{align}
which indicates that $\det [H(\boldsymbol{k})q^{\dagger}(\boldsymbol{k})]$ is  real. It follows that
\begin{align}
    \int^{2\pi}_{0} dk_i \frac{\partial}{\partial k_i} \log\det [H(\boldsymbol{k})q^{\dagger}(\boldsymbol{k})]
    =0,
\end{align}
because the integrant $\log \det [H(\boldsymbol{k})q^{\dagger}(\boldsymbol{k})]$ is 
real and periodic in the Brillouin zone.
Eventually, we obtain
\begin{align}\label{eq:q_H_winding_same}
&\int^{2\pi}_{0} dk_i \frac{\partial}{\partial k_i} \log\det [H(\boldsymbol{k})]
=- \int^{2\pi}_{0} dk_i \frac{\partial}{\partial k_i} \log\det [q^{\dagger}(\boldsymbol{k})] \nonumber \\
=&\int^{2\pi}_{0} dk_i \frac{\partial}{\partial k_i} \log\det [q(\boldsymbol{k})],
 \end{align}
where we have used $\log \det q(\boldsymbol{k})=-\log \det q^{\dagger}(\boldsymbol{k})$ because of the unitarity of $q(\boldsymbol{k})$. 
From this equation, it is clear that $\nu_{{i}}(k_{\perp})$ is the same as $W_{{i}}(k_{\perp})$. 

\begin{figure}
\includegraphics[width=0.8\columnwidth]{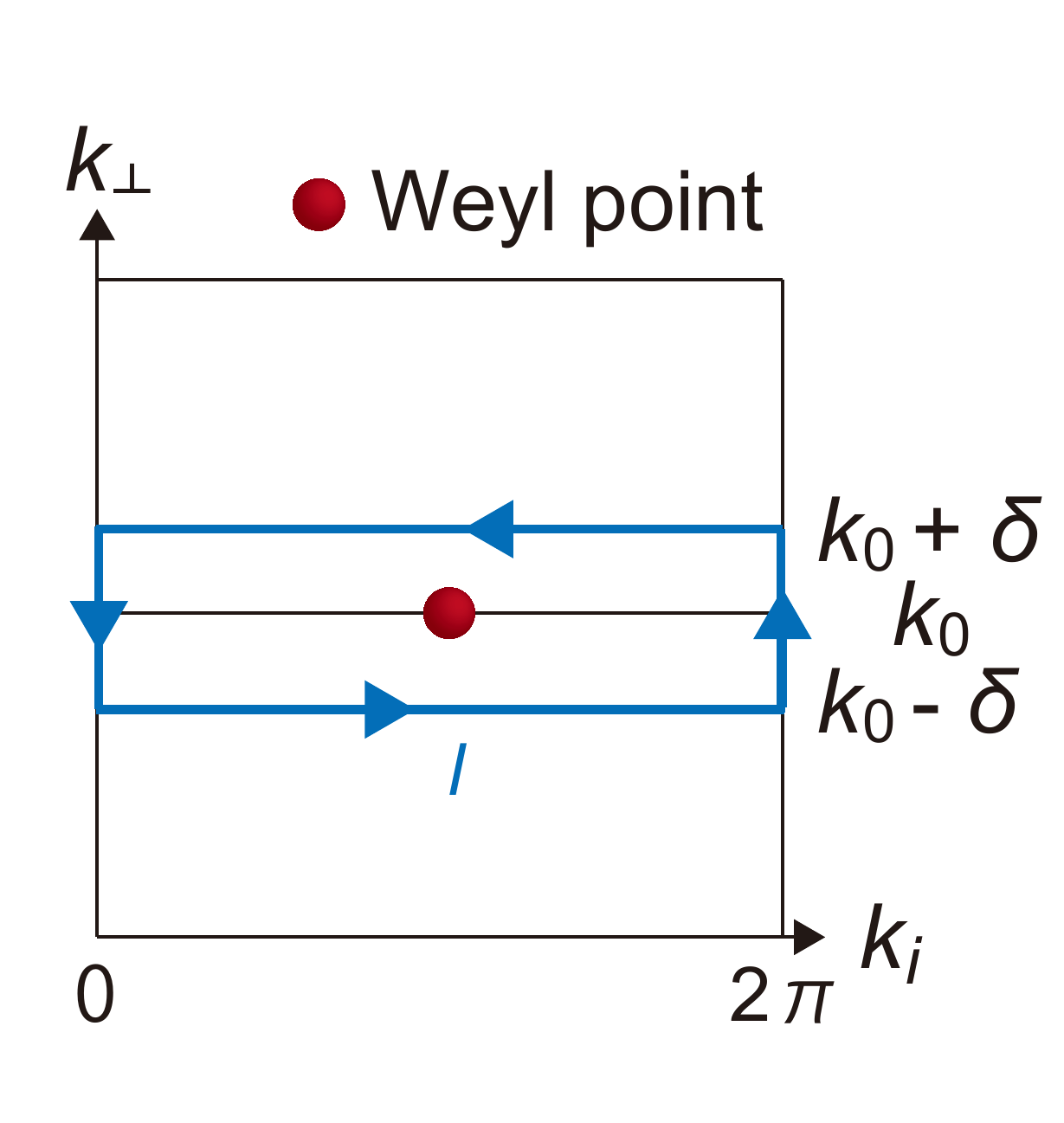}
\caption{An integral path $l$ for the winding number encircling a Weyl point at $k_{\perp}=k_0$.   
} 
\label{fig:winding_path}
\end{figure}

\subsection{Winding number and Chirality of Weyl points}
Here we review a relationship between a winding number and 2D Weyl points. 
In the presence of chiral symmetry, a 2D Weyl point at zero energy can have a nonzero chirality related to the winding number.
When a 2D Weyl point is located at $\boldsymbol{k}_0$ in the Brillouin zone, 
the chirality is given by \cite{Matsuura_2013,  PhysRevB.106.045126}
\begin{align}\label{sup_eq_winding_number}
\nu(\boldsymbol{k}_0)
=\frac{1}{2\pi i}\oint_{l(\boldsymbol{k}_0)} d\boldsymbol{k} \cdot \partial_{\boldsymbol{k}} \log \det [q(\boldsymbol{k})],
\end{align}
where $l$ is a loop encircling the Weyl point.
Moreover, when time-reversal symmetry $\tilde{\mathcal{T}}$ is present, 
another Weyl point can appear at $-\boldsymbol{k}_0$ if $\boldsymbol{k}_0$ is not a time-reversal-invariant momentum.
The chirality can be determined by whether the time-reversal and the chiral operators commute or anticommute.
We write the relation as
 \begin{gather}
     \tilde{\mathcal{T}}\Gamma = c_{\tilde{\mathcal{T}}} \Gamma \tilde{\mathcal{T}},\\ 
     c_{\tilde{\mathcal{T}}}=\pm 1.
 \end{gather}
Then, winding numbers encircling each of the 2D Weyl points at $\boldsymbol{k}_0$ and $-\boldsymbol{k}_0$ are related via the relation \cite{PhysRevB.106.045126}
\begin{equation}\label{eq:trs_winding_conclusion}
    \nu (-\boldsymbol{k}_0)=-c_{\tilde{T}} \nu (\boldsymbol{k}_0).
\end{equation}

We also show how the winding number in Eq.~(\ref{smeq:line_winding}) is related to the chirality.
Let $k_{\perp}$ be the wave vector perpendicular to $k_{i}$.
We choose as $l$ the integral path that encircles only one Weyl point at $k_{\perp}=k_0$, as shown in Fig.~\ref{fig:winding_path}.
Then, since $\tilde{H}(\boldsymbol{k}+\boldsymbol{G})=\tilde{H}(\boldsymbol{k})$ because of the gauge choice, 
the winding number is written as 
\begin{gather}
\nu(\boldsymbol{k}_0)=\nu_{i}(k_0-\delta)-\nu_{i}(k_0+\delta),
\end{gather}
where $\delta$ is a real and positive value. 
Because the winding number $\nu(\boldsymbol{k}_0)$ is nonzero, either  $\nu_{{i}}(k_0+\delta)$ or $\nu_{{i}}(k_0-\delta)$ is nonzero.
Thus, the Weyl point leads to the nonzero winding number $W_{{i}}(k_{\perp})$
since $\nu(k_{\perp}) = W_{{i}}(k_{\perp})$.

We can generalize the idea to 3D systems.
3D chiral-symmetric systems can realize nodal lines due to band crossing instead of Weyl points
because band degeneracy forms a gapless line in the 3D Brillouin zone \cite{RevModPhys.88.035005}.
The nodal line can be characterized by the same winding number in Eq.~(\ref{sup_eq_winding_number}).
When we choose a loop linking with the nodal line as the integral path,
the winding number is nonzero.
By deforming the loop to a line in the same way as 2D,
the system with the nodal line has a nonzero winding number in Eq.(\ref{smeq:line_winding}), which is equivalent to in Eq.~(\ref{winding}) in the main text.

	\section{Screw symmetry with $c_g=1$ in 2D class AII$^{\dagger}$}\label{appendix:cg=1}
	
	\begin{figure}
	\includegraphics[width=1.\columnwidth]{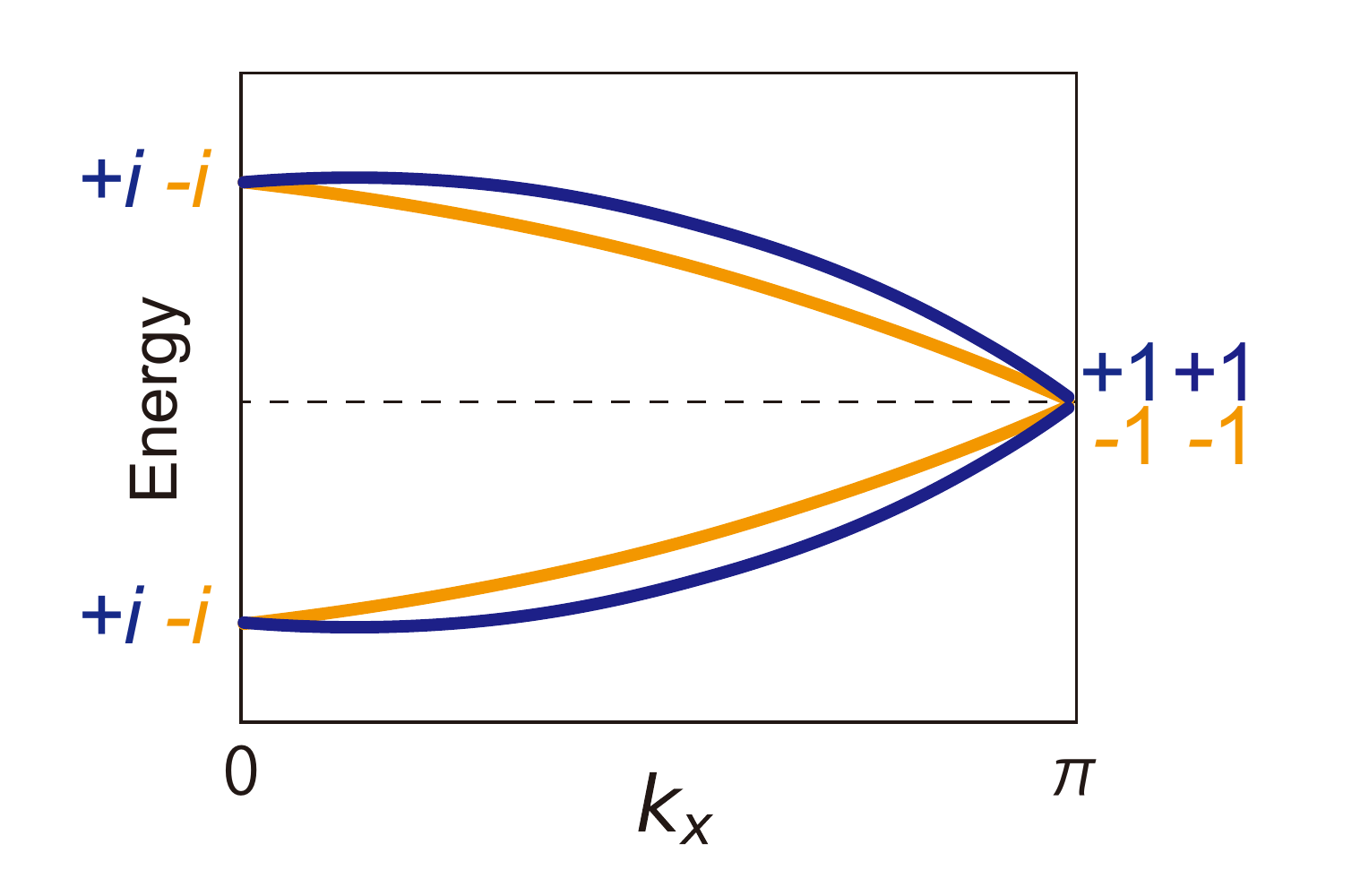}
	\caption{
		Schematic band structures of a doubled Hermitian Hamiltonian with  $\{ C_{2x} | \hat{\boldsymbol{x}}/2 \}$ symmetry and time-reversal symmetry along the screw-invariant line. $\Gamma$ and $\tilde{U}^{\boldsymbol{k}}_{g}$ satisfy commutation relation ($c_g=1$).
	} \label{fig:concept_AIIdagger_cg=1}
\end{figure}

In this appendix, we discuss screw symmetry $g=\{ C_{2x} | \hat{\boldsymbol{x}}/2 \}$ with $c_g=1$ in Eq.~(\ref{eq:symmetry_non_hermitian_ab}) in class AII$^{\dagger}$. 
We apply the symmetry condition in Eq.~(\ref{eq:symmetry_non_hermitian_ab}) to the screw symmetry 
$\tilde{U}^{\boldsymbol{k}}_{g}\tilde{H}(\boldsymbol{k})(\tilde{U}^{\boldsymbol{k}}_{g})^{-1}
		=\tilde{H}(C_{2x}\boldsymbol{k})$.
Since $\{ C_{2x} | \hat{\boldsymbol{x}}/2 \}{^2}=\{ C_{2x}^2 | \hat{\boldsymbol{x}} \}$,
it follows that $\tilde{U}^{C_{2x} \boldsymbol{k}}_{g}\tilde{U}^{\boldsymbol{k}}_{g}
	= e^{-ik_x}$ $(-e^{-ik_x})$ for $C_{2x}^{2}=1(-1)$. 
Therefore, we obtain
	\begin{align}
		& c_sA^{\boldsymbol{k}}_{g} =(c_s e^{ik_{x}} A^{C_{2x}\boldsymbol{k}}_{g})^{\dagger} \ (c_{g} =1 ),  \nonumber \\
		&c_s B^{\boldsymbol{k}}_{g} =(c_s e^{ik_{x}}
		B^{C_{2x}\boldsymbol{k}}_{g})^{\dagger}\  (c_{g} =1 ), \label{supeq:generalized_screw_B_A_relation_classA_c=1} 
	\end{align}
	where $c_s=1(i)$ for $C_{2x}^{2}=1(-1)$, and  $A^{\boldsymbol{k}}_{g}$ and $B^{\boldsymbol{k}}_{g}$ are unitary matrices in Eq.~(\ref{sup:symmetry_rep}). 
	When Eq.~(\ref{eq:symmetry_non_hermitian_ab}) and  Eq.~(\ref{supeq:generalized_screw_B_A_relation_classA_c=1}) hold, the doubled Hermitian Hamiltonian has the screw symmetry.
	
    Because the chiral operator commutes with the symmetry representation $\tilde{U}_g^{\boldsymbol{k}}$ here, a pair of eigenstates at energy $\mathcal{E}$ and $-\mathcal{E}$ have the same eigenvalues of the screw symmetry \cite{comment2}.
    At the same time, the time-reversal symmetry gives rise to Kramer's degeneracy of two states with the same screw eigenvalues.
    Therefore, the screw symmetry with $c_g=1$ leads to a four-fold degeneracy at $k_x=\pi$ (Fig.~\ref{fig:concept_AIIdagger_cg=1}), and this case is beyond our scope.

\section{Mechanism of symmetry-enforced NHSEs in 2D class A}\label{sm:symmetry_condition}

 \begin{figure}
\includegraphics[width=1.\columnwidth]{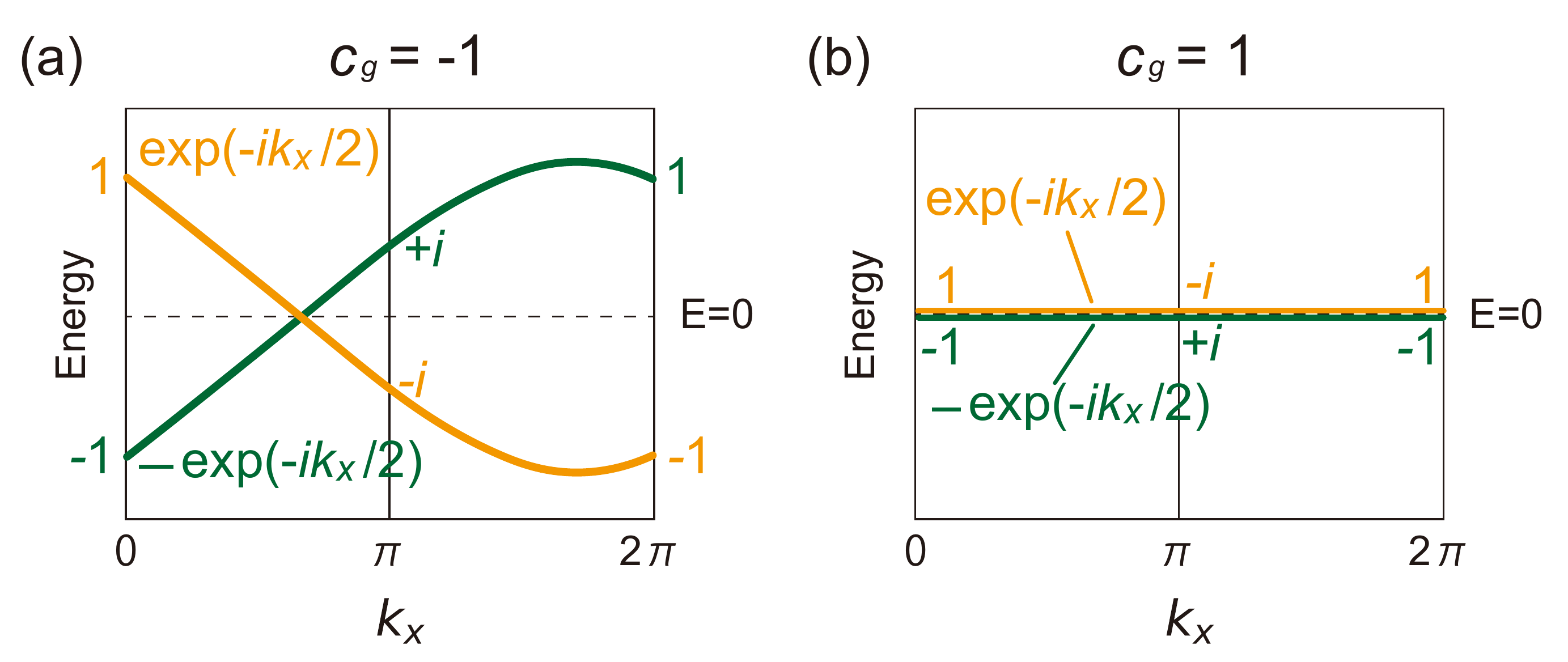}
\caption{Weyl point and gapless line in a 2D doubled Hermitian Hamiltonian with screw symmetry in class AIII.
The yellow (green) lines represent energy eigenvalues with the screw symmetry eigenvalue $+e^{-ik_x/2} (-e^{ik_x/2})$.
(a) and (b) show energy bands along a screw-invariant line in the cases where the chiral and the screw symmetries satisfy the anticommutation relation ($c_g=-1$) and commutation relation ($c_g=1$), respectively. 
} 
\label{fig:symmetry_enforced_gapless_classA}
\end{figure}

We henceforth show that screw symmetry with $c_g=-1$ yields the nonsymmorphic symmetry-enforced NHSE.
We assume that the matrix size of $H(\bm{k})$ is $2M+1$ with a nonnegative integer $M$.
When $H(\bm{k})$ satisfies Eq.~(\ref{eq:generalized_screw_symmetry}), 
the doubled Hermitian Hamiltonian $\tilde{H}(\bm{k})$ has the screw rotation symmetry.
Eigenstates of $\tilde{H}(\bm{k})$ can be labeled by the symmetry eigenvalues $\pm c_se^{-i\frac{k_x}{2}}$
on the screw-invariant lines.
By replacing $k_x$ with $k_x+2\pi$,
	the eigenvalues change from $\pm c_s e^{-i\frac{k_x}{2}}$ to $\mp c_se^{-i\frac{k_x}{2}}$.
	Therefore, pairs of bands with opposite eigenvalues must intersect an odd number of times \cite{PhysRevB.59.5998} from $k_x=0$ to $k_x=2\pi$ on the screw-invariant line, which enforces 2D Weyl points \cite{PhysRevLett.115.126803}.

We consider the effect of chiral symmetry on the band structures of the doubled Hermitian Hamiltonian.
Because of the chiral symmetry $\Gamma$, an eigenstate with the energy $\mathcal{E}$ appears with an eigenstate with $-\mathcal{E}$ in pairs.
When $\Gamma$ and $\tilde{U}^{\boldsymbol{k}}_{g}$ satisfy anticommutation relation (i.e. $c_g=-1$),
a pair of the eigenstates has the screw eigenvalues with opposite signs [Fig.~\ref{fig:symmetry_enforced_gapless_classA}(a)].
The two eigenstates with opposite screw symmetry eigenvalues intersect an odd number of times along the screw-invariant line. The crossing points are 2D Weyl points.
By assumption about the matrix size, the Weyl points appear at zero energy, which gives a nonzero winding number.

 \begin{figure}[b]
\includegraphics[width=1.\columnwidth]{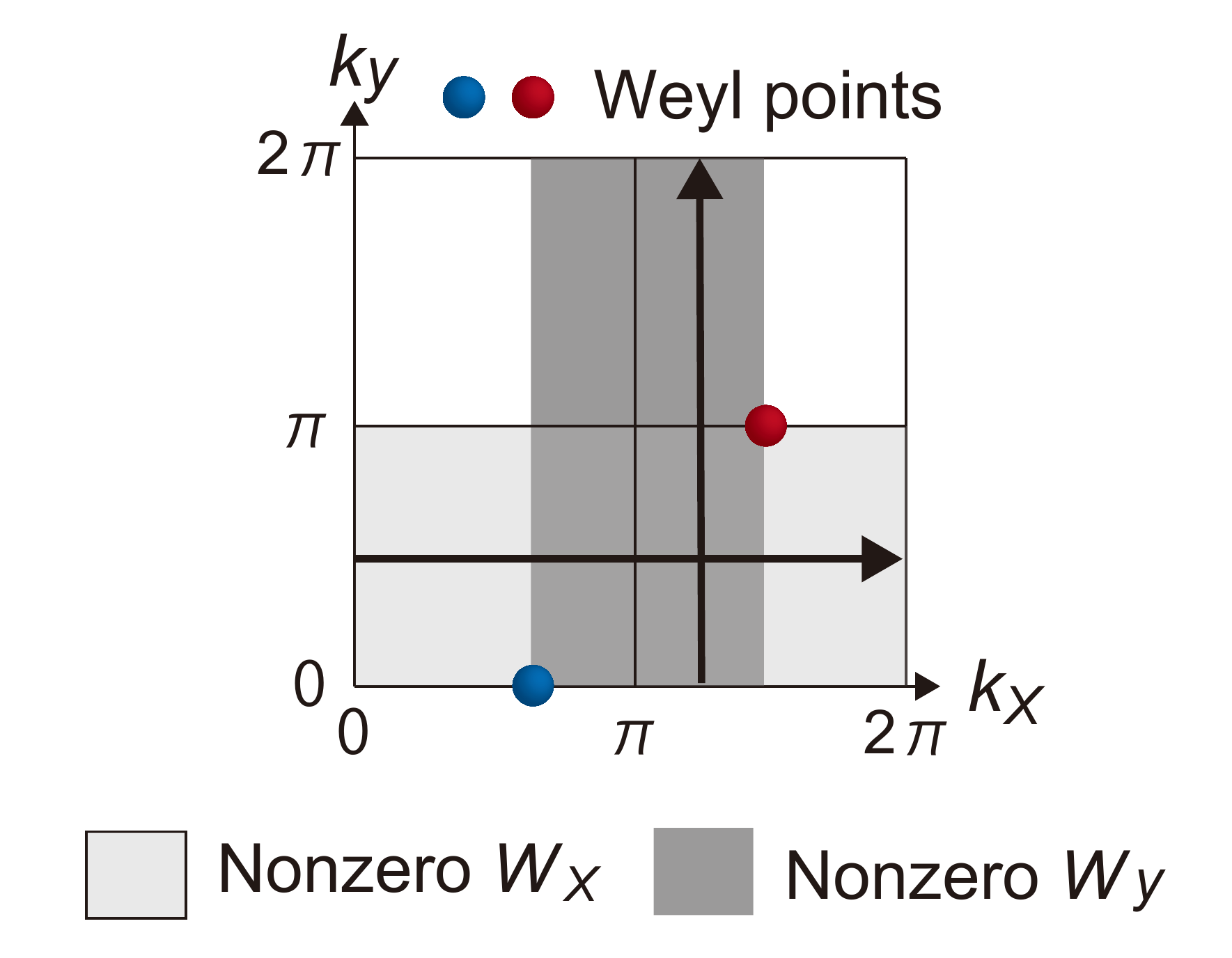}
\caption{Weyl points and regions with the nonzero winding number $W_{i} (k_{\perp})$ ($i=x,y$). The arrows indicate the direction of the integration of $W_{i} (k_{\perp})$. 
			The red and the blue points are Weyl points, and the color difference represents their opposite chirality. 
} 
\label{fig:class_A_weyl_point}
\end{figure}

In contrast, when $c_g=1$, the pair of states with the energy $\mathcal{E}$ and $-\mathcal{E}$ have the same screw symmetry  eigenvalues [Fig.~\ref{fig:symmetry_enforced_gapless_classA}(b)]. Therefore, the screw symmetry with $c_g=1$ leads to gap-closing lines at zero energy.
Hence, the 2D Weyl points do not appear in this case, and we do not consider this case in our paper.

	When the doubled Hermitian Hamiltonian $\tilde{H}(\boldsymbol{k})$ has a nonzero winding number, the non-Hermitian Hamiltonian $H(\boldsymbol{k})$ also has nonzero $W_{i}(k_{\perp})$, as discussed above. 
	Thus, corresponding to the 2D Weyl points enforced by the screw symmetry, the non-Hermitian Hamiltonian $H(\boldsymbol{k})$ has a nonzero winding number [Fig.~\ref{fig:class_A_weyl_point}].
    Also, because the positions of the Weyl points are fixed on the screw-invariant lines $(k_y=0, \pi)$, there is always at least one direction where the winding number is nonzero regardless of the positions of the Weyl points.
	As a result, a NHSE along some direction is also enforced by the screw symmetry of $\tilde{H}(\boldsymbol{k})$.
	Namely, because a winding number is necessarily nonzero in the Brillouin zone thanks to the symmetry-enforced Weyl points,
	the NHSE is robust as long as $H(\boldsymbol{k})$ satisfies Eq.~(\ref{eq:generalized_screw_symmetry}).

Similarly to class AII$^{\dagger}$, we can find anisotropy of winding numbers for the symmetry-enforced NHSE in class A.
Since two Weyl points are enforced to appear on the different screw-invariant lines,
we can always find a path for a nonzero $W_x(k_y)$ in the Brillouin zone,
as shown in Fig.~\ref{fig:class_A_weyl_point}.
Thus, the symmetry-enforced NHSE under the OBC in the $x$ direction is more robust than that under the OBC in the $y$ direction.

 \begin{figure}
		\includegraphics[width=1.\columnwidth]{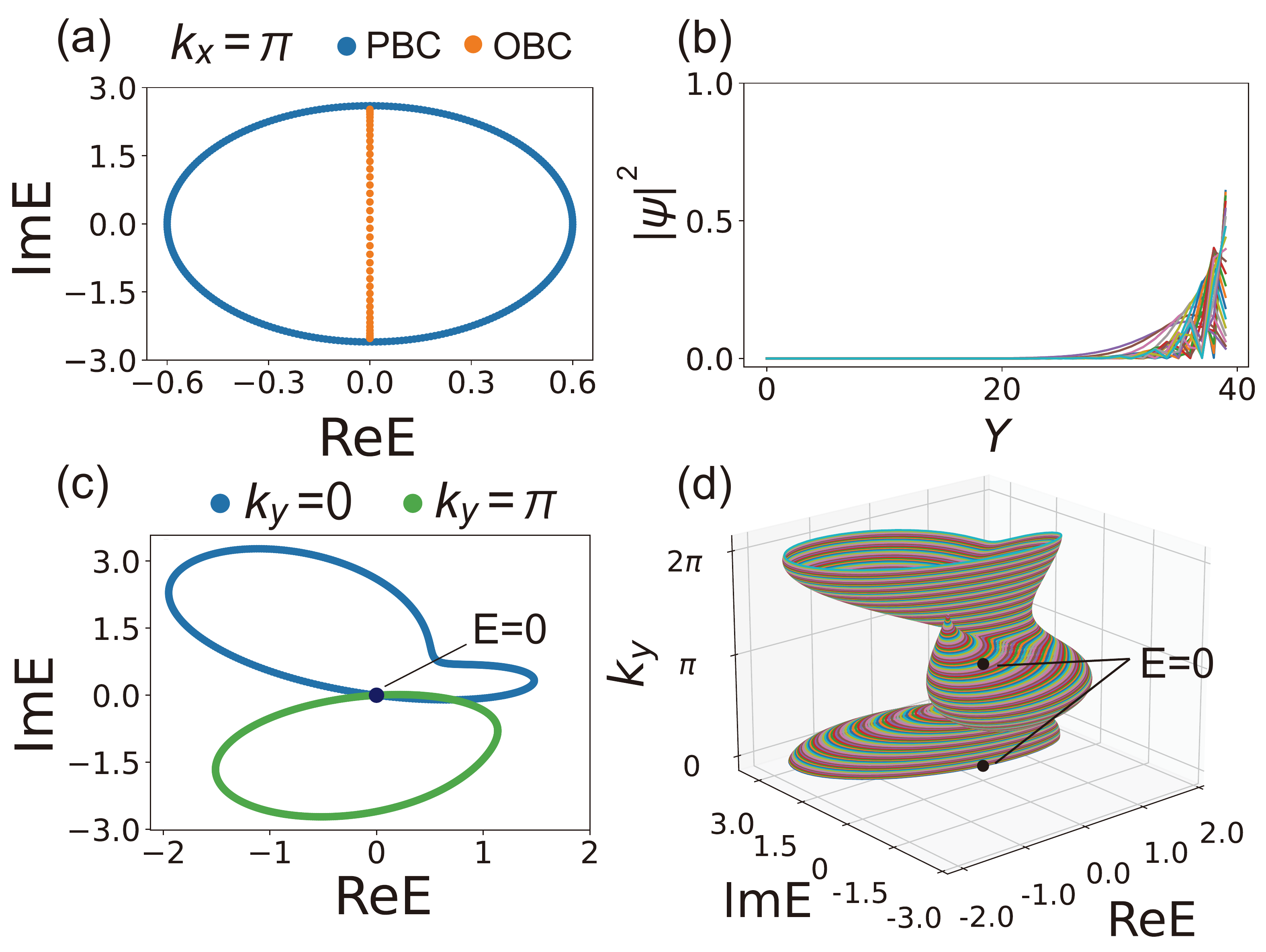}
\caption{Energy spectra and eigenstates of $H^{\rm A}_{\rm 2D}(\boldsymbol{k})$ with the parameters $t_{1}=0.4$, $t_{2}=0.1$, $t_{3}=0.1$, $t_{4}=0.8$, and $t_{5}=0.5$. 
			We take the PBC in the $x$ direction for all the calculations. 
			(a) The energy spectra on the $k_x=\pi$ line under the PBC and the OBC in the $y$ direction. (b) The spatial distributions of all the right eigenstates $\psi _n$ on the $k_x=\pi$ line. 
			(c) The energy spectra on the $k_y=0$ and $k_y=\pi$ lines under the full PBC.
			(d) The energy spectra in $E$-$k_y$ space. 
		} 
		\label{fig:classA_model}
	\end{figure}

\begin{figure*}
\includegraphics[width=2.\columnwidth]{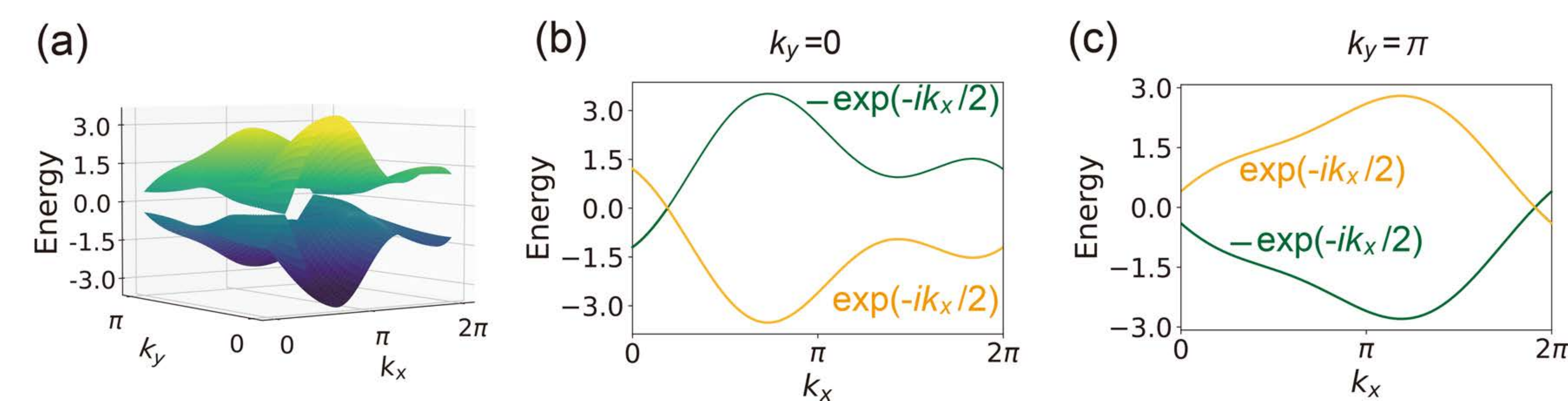}
\caption{(a) The band structures of $\tilde{H}^{\rm AIII}_{\rm 2D}(\boldsymbol{k})$. (b,c) The energy bands of $\tilde{H}^{\rm AIII}_{\rm 2D}(\boldsymbol{k})$ along the screw-invariant line (b) at $k_{y}=0$  and (c) at $k_{y}=\pi$.
Here we choose the parameters $t_{1}=0.4$, $t_{2}=0.1$, $t_{3}=0.1$, $t_{4}=0.8$, and $t_{5}=0.5$.} 
\label{fig:s2_extended_model}
\end{figure*}

\subsection{Model calculation in class A}
We introduce a tight-binding model to demonstrate a nonsymmorphic-symmetry-enforced NHSE in class A.
The Hamiltonian on the square lattice is given by
\begin{align}
    H^{\rm A}_{\rm 2D}
    =&\sum_{\boldsymbol{R}} \sum_{\boldsymbol{r}_{x}= -\boldsymbol{e}_{x}, 2\boldsymbol{e}_{x}}\biggl( 
	t_{1}\ket{\boldsymbol{R}+\boldsymbol{r}_{x}}\bra{\boldsymbol{R}}\nonumber \\
	&+t_{2}\ket{\boldsymbol{R}+\boldsymbol{r}_{x}+\boldsymbol{e}_{y}}\bra{\boldsymbol{R}}
    +t_{3}\ket{\boldsymbol{R}+\boldsymbol{r}_{x}}\bra{\boldsymbol{R}+\boldsymbol{e}_{y}} \biggr)\nonumber \\
	&+\sum_{\boldsymbol{R}} \sum_{\boldsymbol{r}_{x}=0, \boldsymbol{e}_{x}}\biggl(it_{4}(-1)^{\boldsymbol{r}_{x}\cdot \boldsymbol{e}_{x}}
    \ket{\boldsymbol{R}+\boldsymbol{r}_{x}+\boldsymbol{e}_{y}}\bra{\boldsymbol{R}} \nonumber \\
    &+it_{5}(-1)^{\boldsymbol{r}_{x}\cdot \boldsymbol{e}_{x}}\ket{\boldsymbol{R}+\boldsymbol{r}_{x}}\bra{\boldsymbol{R}+\boldsymbol{e}_{y}}
	\biggr).
\end{align}
By using the Fourier transformation under the PBC, we obtain the Bloch Hamiltonian
\begin{align}
		H^{\rm A}_{\rm 2D}(\boldsymbol{k})=&(e^{ik_{x}}+e^{-2ik_{x}})(t_{1}+t_{2}e^{-ik_{y}}+t_{3}e^{i k_{y}}) \nonumber \\
		&+i(1-e^{-ik_{x}})(t_{4}e^{-ik_{y}}+t_{5}e^{ik_{y}}).
\end{align}
	The Hamiltonian $H^{\rm A}_{\rm 2D}(\boldsymbol{k})$ satisfies the condition in Eq. (\ref{eq:generalized_screw_symmetry}) with $c_s=1$ and  $A^{\boldsymbol{k}}_{g}=1$. Figure \ref{fig:classA_model}(a) shows energy spectra of $H^{\rm A}_{\rm 2D}(\boldsymbol{k})$ on the $k_{x}=\pi$ line under both the full PBC and the OBC in the $y$ direction. 
	Figure \ref{fig:classA_model}(b) shows spatial distributions for all the eigenstates,
	and we can see that they are localized at the right edge. 
	These results indicate that a NHSE occurs in the model.  
	Moreover, the point gap closes at $E=0$ on the $k_y=0$ and $k_y=\pi$ lines under the full PBC [Figs.~\ref{fig:classA_model}(c) and \ref{fig:classA_model}(d)]. 

Next, we investigate a doubled Hermitian Hamiltonian $\tilde{H}^{\rm AIII}_{\rm 2D}(\boldsymbol{k})$ obtained from $H^{\rm A}_{\rm 2D}(\boldsymbol{k})$.
The doubled Hermitian Hamiltonian can be written as 
\begin{align}
\tilde{H}^{\rm AIII}_{\rm 2D}(\boldsymbol{k})
=&\begin{pmatrix}
 & H^{\rm A}_{\rm 2D}(\boldsymbol{k}) \\
H^{\rm A}_{\rm 2D}(\boldsymbol{k})^{\dagger} &
\end{pmatrix} \\
=&f_{x}(\boldsymbol{k})\sigma_{x}+f_{y}(\boldsymbol{k})\sigma_{y},
\end{align}
with 
\begin{align}
    f_{x}(\boldsymbol{k})=&(\cos k_{x}+ \cos 2k_x)(t_{1}+t_{2}\cos k_{y}+t_{3}\cos k_{y})\nonumber \\
    &+(-\sin k_{x}+\sin 2k_x)(-t_{2}\sin k_{y}+t_{3}\sin k_{y})\nonumber \\
	&+(t_{4}-t_{5})(1-\cos k_{x})\sin k_{y}\nonumber \\
	&-(t_{4}+t_{5}) \sin k_{x} \cos k_{y}, \nonumber \\
	f_{y}(\boldsymbol{k})=&-(\cos k_{x}+\cos 2k_{x})(-t_{2}\sin k_{y}+t_{3}\sin k_{y})\nonumber \\
	&-(\sin k_{x}-\sin 2k_x)(t_{1}+t_{2}\cos k_{y}+t_{3}\cos k_{y})\nonumber \\
	&-(t_{4}+t_{5})(1-\cos k_{x})\cos k_{y} \nonumber \\
	&+(-t_{4}+t_{5}) \sin k_{x}\sin k_{y}.
\end{align}
The doubled Hermitian Hamiltonian has chiral symmetry represented by $\Gamma \tilde{H}^{\rm AIII}_{\rm 2D}(\boldsymbol{k})\Gamma^{-1}=-\tilde{H}^{\rm AIII}_{\rm 2D}(\boldsymbol{k})$ with $\Gamma=\sigma_{z}$. 
The screw $g=\{ C_{2x} | \hat{\boldsymbol{x}}/2 \}$ symmetry of the doubled Hermitian Hamiltonian is represented by 
\begin{align}
  	\tilde{U}^{\boldsymbol{k}}_{g} \tilde{H}^{\rm AIII}_{\rm 2D}(\boldsymbol{k}) (\tilde{U}^{\boldsymbol{k}}_{g} )^{-1}=  \tilde{H}^{\rm AIII}_{\rm 2D}(C_{2x}\boldsymbol{k}),
\end{align}
with
\begin{align}
	\tilde{U}^{\boldsymbol{k}}_{g}=\begin{pmatrix}
	& e^{-ik_{x}} \\
	1 &
	\end{pmatrix}.
\end{align}
Because $\tilde{U}^{C_{2x}\boldsymbol{k}}_{g}\tilde{U}^{\boldsymbol{k}}_{g} =e^{-ik_{x}}$ in the model, the screw symmetry eigenvalues are $\pm e^{-ik_{x}/2}$.
Moreover, $\tilde{U}^{\boldsymbol{k}}_{g}$ and $\Gamma$ satisfy the anticommutation relation $\tilde{U}^{\boldsymbol{k}}_{g} \Gamma = -\Gamma \tilde{U}^{\boldsymbol{k}}_{g}$
, which corresponds to $c_g=-1$.

Figure \ref{fig:s2_extended_model}(a) shows the band structure of the doubled Hermitian Hamiltonian $\tilde{H}^{\rm AIII}_{\rm 2D}(\boldsymbol{k})$,
where we can find two Weyl points at zero energy. 
Figures~\ref{fig:s2_extended_model}(b) and \ref{fig:s2_extended_model}(c) show the band structures along the screw-invariant line at $k_{y}=0$ and $k_{y}=\pi$. 
We can see that the two eigenstates with screw eigenvalues with opposite signs switch, leading to the Weyl points. 
Since a winding number is nonzero between the Weyl points, 
the NHSE necessarily occurs. 

\section{Mechanisms of symmetry-enforced NHSEs in 2D class AII}\label{appendx:classAII}
In this appendix, we study symmetry-enforced NHSEs in 2D class AII.
We clarify the relationship between NHSEs and 2D Weyl points enforced by the screw symmetry when the screw operator commutes with the time-reversal operator.

First, we discuss symmetry condition of a non-Hermitian Hamiltonian in class AII \cite{PhysRevX.9.041015} for the doubled Hermitian Hamiltonian to have time-reversal symmetry.
Let us assume that a non-Hermitian Hamiltonian $H(\boldsymbol{k})$ in class AII satisfies Eq.~(\ref{eq:generalized_screw_symmetry}).
The non-Hermitian Hamiltonian $H(\boldsymbol{k})$ has time-reversal symmetry  
\begin{align}
	\mathcal{T}H^{*}(\boldsymbol{k})\mathcal{T}^{-1}=H(-\boldsymbol{k}),
\end{align}
where $\mathcal{T}$ is a unitary matrix and satisfies 
\begin{align}
	 \mathcal{T} \mathcal{T}^{*}=-1.
\end{align}
The doubled Hermitian Hamiltonian $\tilde{H}(\boldsymbol{k})$ has time-reversal symmetry given by
\begin{align}
	\tilde{\mathcal{T}}\tilde{H}(\boldsymbol{k})\tilde{\mathcal{T}}^{-1}=\tilde{H}(-\boldsymbol{k}),
\end{align} 
where $\tilde{\mathcal{T}}$ is expressed as
\begin{align}
	 \tilde{\mathcal{T}}=\begin{pmatrix}
	\mathcal{T} & \\
	&\mathcal{T}
	\end{pmatrix}K,
\end{align} 
with $\tilde{\mathcal{T}}^{2}=-1$.
The doubled Hermitian Hamiltonian $\tilde{H}(\boldsymbol{k})$ has particle-hole symmetry given by $\tilde{\mathcal{P}}=\tilde{\mathcal{T}}^{-1}\Gamma$,
and therefore $\tilde{\mathcal{P}}^2=-1$ due to $\Gamma \tilde{\mathcal{T}} = \tilde{\mathcal{T}}\Gamma $ (i.e. $c_{\tilde{\mathcal{T}}}=1$). 
Thus, $\tilde{H}(\boldsymbol{k})$ belongs to class CII.
We also impose the condition that $g=\{ C_{2x} | \hat{\boldsymbol{x}}/2 \}$ and $\tilde{\mathcal{T}}$ commute in order to specify their relation.
Then, because $\tilde{\mathcal{T}} \tilde{U}^{\boldsymbol{k}}_{g}  = \tilde{U}^{-\boldsymbol{k}}_{g}   \tilde{\mathcal{T}}$, we obtain
\begin{align}
	\mathcal{T} (A^{\boldsymbol{k}}_{g})^{*}
	=A^{-\boldsymbol{k}}_{g} \mathcal{T}, \label{eq:ag_t_classAII} \\ 
    \mathcal{T} (B^{\boldsymbol{k}}_{g})^{*}=B^{-\boldsymbol{k}}_{g} \mathcal{T}.
\end{align} 
By using the unitary matrix $U_{g}^{\boldsymbol{k}}=i e^{i\frac{k_x}{2}}A^{\boldsymbol{k}}_{g}$, we can rewrite Eq.~(\ref{eq:ag_t_classAII}) as 
\begin{equation}
    \mathcal{T} (U_{g}^{\boldsymbol{k}})^*= -U_{g}^{-\boldsymbol{k}}  \mathcal{T}.
\end{equation}

\begin{figure}
\includegraphics[width=1.\columnwidth]{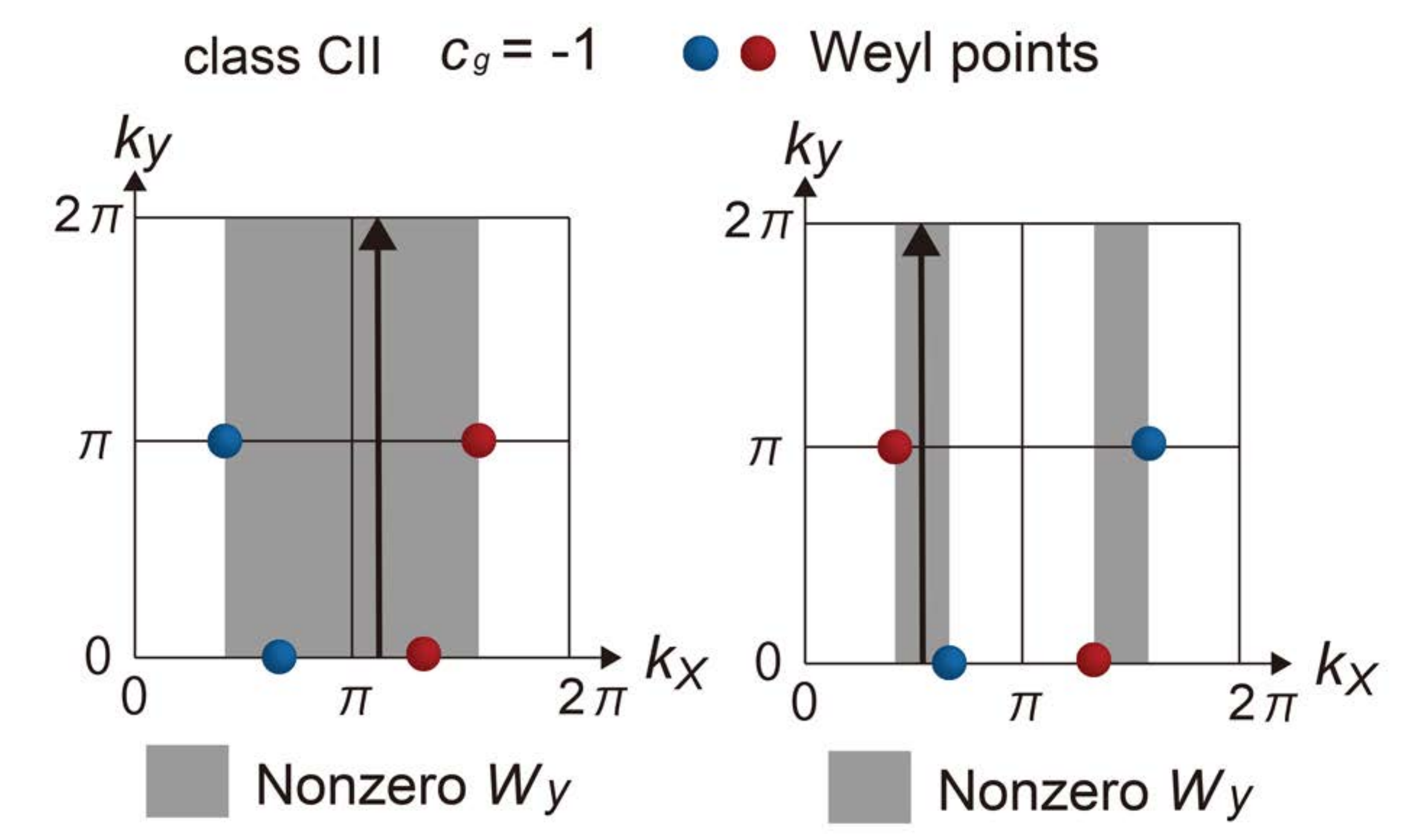}
\caption{
Illustrative examples of the positions of the Weyl points with $c_g =-1$ in class CII and the regions with the nonzero winding number $W_{y} (k_{x})$ in the Brillouin zone.
When a non-Hermitian Hamiltonian belongs to class AII,
the class of the doubled Hermitian Hamiltonian is CII.
The arrows indicate the direction of integration of $W_{y} (k_{x})$. The colors (red and blue) of the points  represent the opposite chirality of the Weyl points. Here, we assume $W_{y} (0)=0$.} 
\label{fig:symmetry_enforced_gapless}
\end{figure}

Similarly to class AII$^{\dagger}$, we consider the case for $c_s=i$ since $\tilde{\mathcal{T}}^{2}=-1$ corresponds to spinful systems with $C_{2x}^2=-1$, and we assume that the non-Hermitian Hamiltonian has a matrix size of $4M+2$.
Thus, we obtain the band structure of the doubled Hermitian Hamiltonian similar to that in class AII$^{\dagger}$ [Fig.~\ref{fig:concept_AIIdagger}(a)]. This band structure enforces the 2D Weyl points on the screw-invariant lines. It follows that the non-Hermitian Hamiltonian $H(\boldsymbol{k})$ has a nonzero winding number  $W_i(k_{\perp})$ between the two 2D Weyl points in the doubled Hermitian Hamiltonian $\tilde{H}(\boldsymbol{k})$. 

Next, we discuss the nonzero winding numbers due to the positions of the Weyl points with $c_g=-1$ in the Brillouin zone. 
	By using Eq.~(\ref{eq:trs_winding_conclusion}) with $c_{\tilde{\mathcal{T}}}=1$,
	we find that the time-reversal pair of Weyl points on the screw-invariant line has opposite chirality.
	Therefore, two types of positions of the 2D Weyl points can be realized, as shown in Fig.~\ref{fig:symmetry_enforced_gapless}. 
	Because of these positions of the Weyl points, the winding number $W_{x}(k_y)$ is zero in Fig.~\ref{fig:symmetry_enforced_gapless}.
	In contrast, the winding number $W_{y}(k_x)$ can be nonzero.

\subsection{Model calculation in class AII}

\begin{figure*}
\includegraphics[width=1.8\columnwidth]{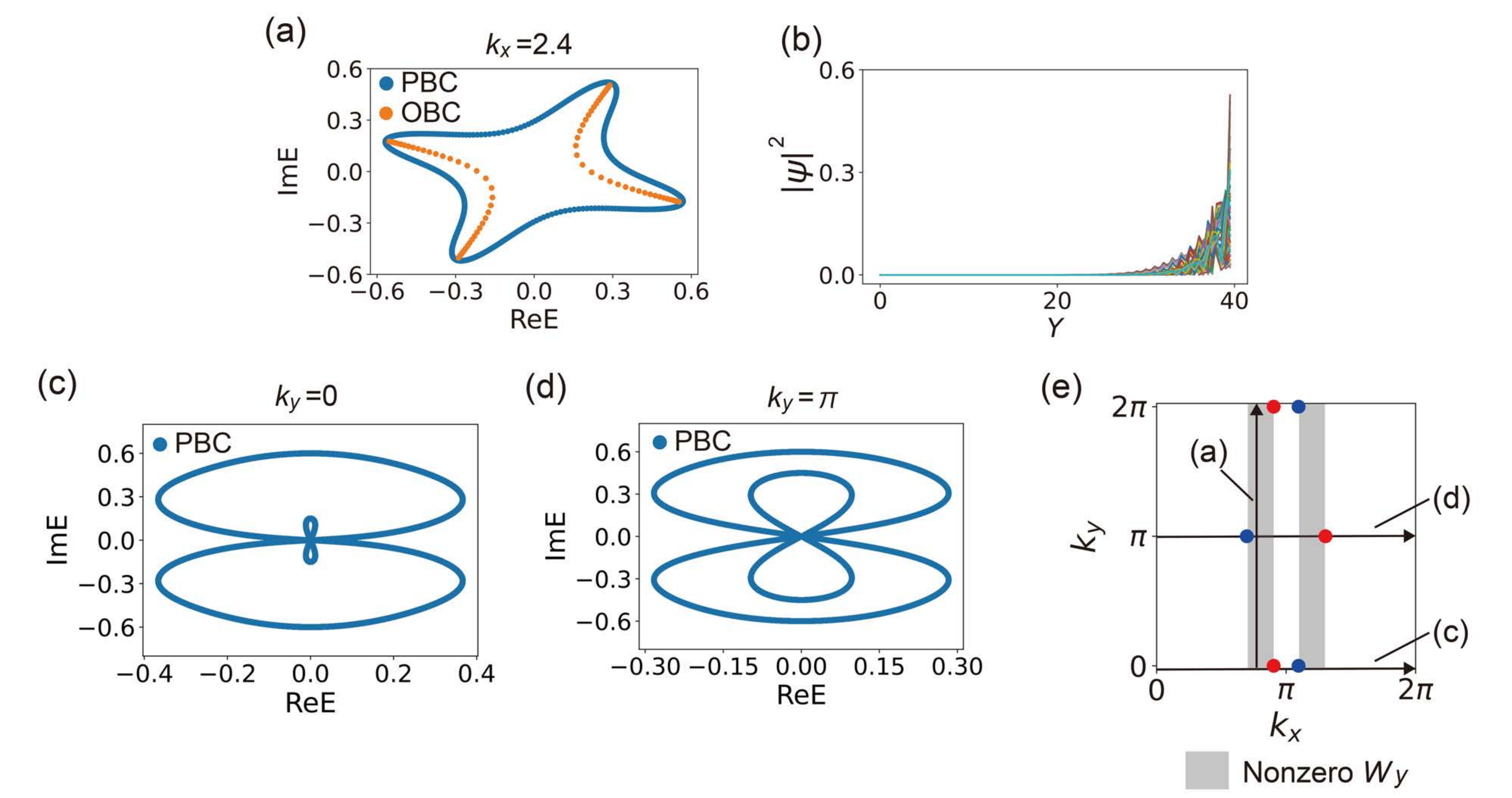}
\caption{Energy spectra of $H^{\rm AII}_{\rm 2D}(\boldsymbol{k})$ with the parameters $t=v=0.15$ and $c=0.05$. (a) The energy spectra under the PBC in the $x$ direction. We apply the PBC in the $y$ direction and the OBC in the $y$ direction. (b) The spatial distributions of $|\psi_{n}|^{2}$ for all the right eigenstates $\psi_{n}$ under the OBC in (a).
The system size in the $y$ direction is 40.
(c), (d)  The energy spectra under the full PBC. The wave vector is fixed to (c) $k_y=0$ and (d) $k_y=\pi$.
(e) The positions of the 2D Weyl point of $H^{\rm CII}_{\rm 2D}(\boldsymbol{k})$ and the regions with the nonzero winding number $W_{y}(k_x)$ in the Brillouin zone.
The arrows indicate the direction of integration of the winding number. The colors of the points represent the opposite chirality of the Weyl points.} 
\label{fig:classAII_NHSE}
\end{figure*}

To show a symmetry-enforced NHSE in class AII,
let us introduce a 2D tight-binding model on a square lattice with the two degrees of freedom $A$ and $B$ in each unit cell. This model is given by 
\begin{align}
	H^{\rm AII}_{\rm 2D}=H^{\rm AII (1)}_{\rm 2D}+H^{\rm AII (2)}_{\rm 2D}+H^{\rm AII (3)}_{\rm 2D},
\end{align}
with
\begin{widetext}
\begin{align}
	H^{\rm AII(1)}_{\rm 2D}=&-it\sum_{\boldsymbol{R}} \sum_{\boldsymbol{r}=0,\boldsymbol{e}_{x}, \boldsymbol{e}_{y},\boldsymbol{e}_{x}+\boldsymbol{e}_{y}} (-1)^{\boldsymbol{r}\cdot(\boldsymbol{e}_{x}+\boldsymbol{e}_{y})}( 
	  \ket{\boldsymbol{R}+\boldsymbol{r},B} \bra{\boldsymbol{R}, A} +  \ket{\boldsymbol{R}+\boldsymbol{r}, A} \bra{ \boldsymbol{R}, B} ) \nonumber \\
	  &-0.5it\sum_{\boldsymbol{R}} \sum_{\boldsymbol{r}=0,\boldsymbol{e}_{x}} (-1)^{\boldsymbol{r}\cdot \boldsymbol{e}_{x}} ( 
	  \ket{\boldsymbol{R}+\boldsymbol{r}-\boldsymbol{e}_{y},B} \bra{\boldsymbol{R}, A} +  \ket{\boldsymbol{R}+\boldsymbol{r}-\boldsymbol{e}_{y},A} \bra{ \boldsymbol{R}, B} ),       
\end{align}
\begin{align}
	H^{\rm AII(2)}_{\rm 2D}=& v \sum_{\boldsymbol{R}} \sum_{\boldsymbol{r}_{x}=0,\boldsymbol{e}_{x}} ( 
	 -2 \ket{\boldsymbol{R}+\boldsymbol{r}_{x}, A} \bra{ \boldsymbol{R}, B} +2 \ket{\boldsymbol{R} +\boldsymbol{r}_{x}, B} \bra{\boldsymbol{R}, A}\nonumber \\ 
	  &+\ket{\boldsymbol{R}+\boldsymbol{r}_{x}-\boldsymbol{e}_{y},A} \bra{ \boldsymbol{R},B}
    -\ket{\boldsymbol{R}+\boldsymbol{r}_{x}-\boldsymbol{e}_{y} ,B}\bra{\boldsymbol{R}, A} -\ket{\boldsymbol{R}+\boldsymbol{r}_{x}+\boldsymbol{e}_{y},A} \bra{\boldsymbol{R},B}
    + \ket{\boldsymbol{R}+\boldsymbol{r}_{x}+\boldsymbol{e}_{y}, B}\bra{\boldsymbol{R}, A}), 
\end{align}
\begin{align}
	  H^{\rm AII(3)}_{\rm 2D}=  -ic \sum_{\boldsymbol{R}} \sum_{\boldsymbol{r}_{y}=-\boldsymbol{e}_{y},\boldsymbol{e}_{y}}& ( 
	     \ket{\boldsymbol{R}-\boldsymbol{e}_{x}+\boldsymbol{r}_{y}, A} \bra{\boldsymbol{R}, A} 
      - \ket{\boldsymbol{R} -\boldsymbol{e}_{x}+\boldsymbol{r}_{y},B} \bra{ \boldsymbol{R}, B})\nonumber \\
	     &- \ket{\boldsymbol{R}+\boldsymbol{e}_{x}+\boldsymbol{r}_{y}, A} \bra{\boldsymbol{R}, A}
      + \ket{\boldsymbol{R} +\boldsymbol{e}_{x}+\boldsymbol{r}_{y}, B} \bra{ \boldsymbol{R}, B} -\ket{\boldsymbol{R}+\boldsymbol{r}_{y}, A} \bra{ \boldsymbol{R}, A} 
      + \ket{\boldsymbol{R} +\boldsymbol{r}_{y},B} \bra{ \boldsymbol{R}, B} \nonumber \\
	     & +\ket{\boldsymbol{R}+2 \boldsymbol{e}_{x}+\boldsymbol{r}_{y},A} \bra{\boldsymbol{R}, A} 
      - \ket{\boldsymbol{R} +2 \boldsymbol{e}_{x}+\boldsymbol{r}_{y},B} \bra{ \boldsymbol{R}, B} ).
\end{align}
\end{widetext}
We obtain the $2\times 2$ Bloch Hamiltonian
\begin{align}
	H^{\rm AII}_{\rm 2D}(\boldsymbol{k})=&-it(1-e^{-ik_{x}})(1-e^{-ik_{y}}+0.5e^{ik_{y}})\sigma_{x}\nonumber \\
	&-2iv(1+e^{-ik_{x}})(1-i \sin k_{y})\sigma_{y}\nonumber \\
	&+4c(1-e^{-ik_{x}})\sin k_{x} \cos k_{y} \sigma_{z},
	\end{align}
where $\sigma_i$ ($i=x,y,z$) corresponds to the two degree of freedom $A$ and $B$. 
The model satisfies the condition for the screw symmetry $g=\{ C_{2x} | \hat{\boldsymbol{x}}/2 \}$ with $c_g=-1$,
which is described as
\begin{gather}\label{aii_model_gss}
iA^{\boldsymbol{k}}_{g} H^{\rm AII}_{\rm 2D}(\boldsymbol{k})= 
\Bigl(ie^{ik_x} A^{C_{2x}\boldsymbol{k}}_{g} H^{\rm AII}_{\rm 2D}(C_{2x}\boldsymbol{k}) \Bigr)^{\dagger}, \\
A^{\boldsymbol{k}}_{g}=-ie^{-ik_y}\sigma_x.
\end{gather}
Correspondingly, we have
\begin{align}
 &B^{\boldsymbol{k}}_{g} =- (e^{ik_x}A^{C_{2x}\boldsymbol{k}}_{g})^{\dagger}=-ie^{-i(k_x+k_y)}\sigma_x. 
\end{align}
Our model also has time-reversal symmetry 
\begin{align}\label{sup_aii_model_TRS}
	\mathcal{T}H^{\rm AII *}_{\rm 2D}(\boldsymbol{k})\mathcal{T}^{-1}=H^{\rm AII}_{\rm 2D}(-\boldsymbol{k}),
\end{align}  
with $\mathcal{T}=i\sigma_{y}$, which corresponds to class AII.
Therefore, the symmetry-enforced NHSE occurs in this model. 

To confirm the symmetry-enforced NHSE,
we calculate energy spectra of $H^{\rm AII}_{\rm 2D}(\boldsymbol{k})$ under the PBC in the $x$ direction.
Figure \ref{fig:classAII_NHSE}(a) shows the energy spectra under the PBC and under the OBC in the $y$ direction. 
Indeed, we can find that the spectrum under the full PBC gives the nonzero $W_y(k_x)$ in the Brillouin zone
because $H^{\rm AII}_{\rm 2D}(\boldsymbol{k})$ satisfies the conditions of the symmetry and the matrix size.
As a result,
we can also find eigenstates localized at the edge under the OBC in the $y$ direction [Fig.~\ref{fig:classAII_NHSE}(b)].
Moreover, we can also see symmetry-enforced gaplessness on the lines $k_y=0$ and $k_y=\pi$,
as shown in Figs~\ref{fig:classAII_NHSE}(c) and \ref{fig:classAII_NHSE}(d).

\begin{figure*}
\includegraphics[width=1.8\columnwidth]{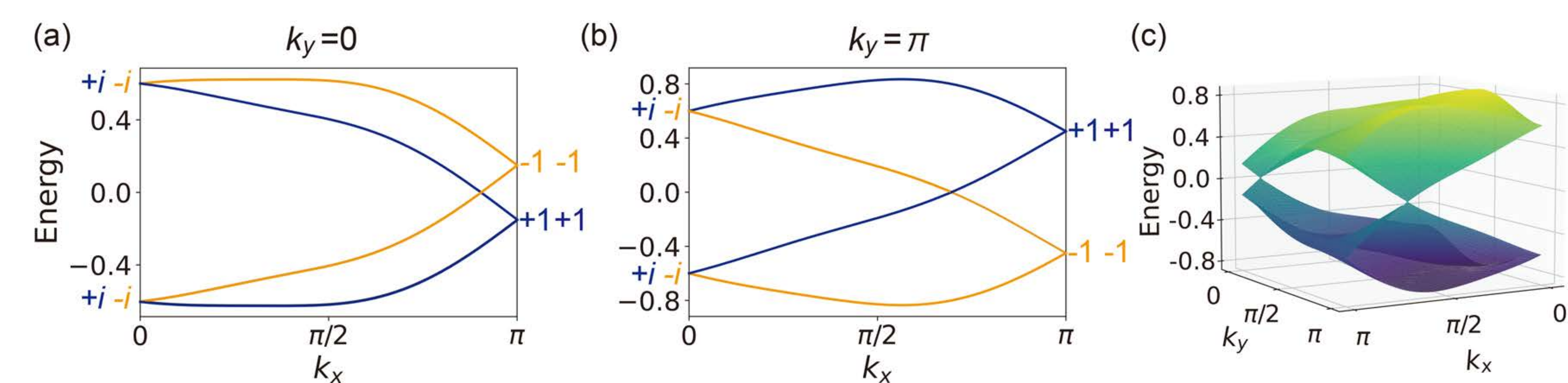}
\caption{
Band structures of $\tilde{H}^{\rm CII}_{\rm 2D}(\boldsymbol{k})$ with the parameters $t=v=0.15$ and $c=0.05$. (a), (b) The energy spectra with the screw $\{ C_{2x} | \hat{\boldsymbol{x}}/2 \}$ symmetry eigenvalues,  $\pm i \exp (-i(k_y+\frac{k_x}{2}))$ along the screw invariant lines with (a) $k_{y}=0$ and (b) $k_{y}=\pi$. (c) The band structures in a quarter of the Brillouin zone.}  
\label{fig:symmetry-enforced_weyl}
\end{figure*}

To see the origin clearly,
we find Weyl points of the doubled Hermitian Hamiltonian
\begin{align}
\tilde{H}^{\rm CII}_{\rm 2D}(\boldsymbol{k})
=&\begin{pmatrix}
 & H^{\rm AII}_{\rm 2D}(\boldsymbol{k}) \\
H^{\rm AII}_{\rm 2D}(\boldsymbol{k})^{\dagger} &
\end{pmatrix} \\
=
    t& \bigl(1.5(1-\cos k_x)\sin k_y \nonumber \\
    &\ \ +\sin k_x (1-0.5 \cos k_y)
    \bigr)\tau_x \otimes \sigma_x \nonumber \\
+&t\bigl((1-\cos k_x)(1-0.5\cos k_y) \nonumber \\
&\ \ -1.5\sin k_x \sin k_y
    \bigr)\tau_y \otimes \sigma_x \nonumber \\
    -&2v\bigl( (1+\cos k_x)\sin k_y + \sin k_x
    \bigr)\tau_x \otimes \sigma_y \nonumber \\
    +&2v \bigl( (1+\cos k_x) -\sin k_x \sin k_y \bigr) \tau_y \otimes \sigma_y \nonumber \\
    +&4c(1-\cos k_x)\sin k_x \cos k_y \tau_x \otimes \sigma_z \nonumber \\
    -&4c \sin^2 k_x \cos k_y \tau_y \otimes \sigma_z.
\end{align}
The doubled Hermitian Hamiltonian has the screw rotation symmetry $\tilde{U}^{\boldsymbol{k}}_{g}\tilde{H}^{\rm CII}_{\rm 2D}(\boldsymbol{k})\tilde{U}^{\boldsymbol{k}\dagger}_{g}=\tilde{H}^{\rm CII}_{\rm 2D}(C_{2x}\boldsymbol{k})$, where $\tilde{U}^{\boldsymbol{k}}_{g}$ is represented as 
\begin{align}
	\tilde{U}^{\boldsymbol{k}}_{g}=-ie^{-ik_{y}}\begin{pmatrix}
&  e^{-ik_{x}} \\
1 & 
\end{pmatrix}\otimes \sigma_{x}. 
\end{align}
Indeed, 2D Weyl points can be found on the screw-invariant lines  [Fig.~\ref{fig:classAII_NHSE}(e)]. 
Thus, the winding number can be nonzero between the Weyl points, as shown in Fig.~\ref{fig:classAII_NHSE}(e).
Figures~\ref{fig:symmetry-enforced_weyl}(a) and \ref{fig:symmetry-enforced_weyl}(b) show the energy spectra and the screw-symmetry eigenvalues along the screw invariant lines ($k_{y}=0, \pi$). In addition, Fig.~\ref{fig:symmetry-enforced_weyl}(c) shows  the band structure in a quarter of the Brillouin zone.
The occupied states with the screw-symmetry eigenvalues require the Weyl point on the screw invariant lines with $k_{y}=0$ and $k_{y}=\pi$. 
Because these Weyl points are enforced by the screw symmetry and time-reversal symmetry, the NHSE in Fig.~\ref{fig:classAII_NHSE}(a) cannot disappear when $H_{\rm 2D}(\boldsymbol{k})$ has the symmetry condition [Eq.~(\ref{aii_model_gss})] and time-reversal symmetry [Eq.~(\ref{sup_aii_model_TRS})]. 

\section{Absence of the symmetry-enforced NHSE in 2D class ${\rm AI}$ and class ${\rm AI}^{\dagger}$}\label{appendx:classAI_and_AIdagger}

We comment on class AI and class ${\rm AI}^{\dagger}$ in 2D systems.
We consider non-Hermitian Hamiltonians satisfying the conditions in Eq.~(\ref{eq:generalized_screw_symmetry}) in class AI and class ${\rm AI}^{\dagger}$.
When a non-Hermitian Hamiltonian belongs to class AI (${\rm AI}^{\dagger}$), the doubled Hermitian Hamiltonian belongs to class BDI (CI). In both classes, the doubled Hermitian Hamiltonian has time-reversal ($\tilde{\mathcal{T}}$) symmetry. 
In a similar way to class A, the screw symmetry of the doubled Hermitian Hamiltonian enforces double degeneracy on the screw-invariant lines ($k_y=0,\pi$) when the matrix size is $4M+2$ \cite{PhysRevLett.115.126803}.
However, because of time-reversal symmetry, the two states form the double degeneracy along the $k_x=\pi$ line because $(\tilde{\mathcal{T}}\{C_{2x}|\hat{\boldsymbol{x}}/2 \})^2=e^{-ik_x}$. Thus, the gap-closing line appears at zero energy instead of 2D Weyl points in both class BDI and class CI, and therefore nonzero winding numbers are not enforced in these cases.
	

\begin{thebibliography}{107}
\bibitem [{com()}]{comment2}%
	\BibitemOpen
	\href@noop {} {\bibinfo  {journal} {The projective factors relate symmetry
			eigenvalues of two eigenstates connected by chiral symmetry. Since doubled
			Hermitian Hamiltonians have chiral symmetry $\Gamma$, an eigenstate
			$\ket{\Psi (\bm{k})}$ with the energy $\mathcal{E}$ appears with the chiral
			partner $\Gamma \ket{\Psi (\bm{k})}$ with $-\mathcal{E}$. The eigenstates
			have symmetry eigenvalues on the high-symmetry points. When
			$\tilde{U}^{\boldsymbol{k}}_{g} \ket{\Psi (\bm{k})}=u_g\ket{\Psi (\bm{k})}$
			where $u_g$ is an eigenvalue of the crystal symmetry $g$, the chiral partner
			satisfies $\tilde{U}^{\boldsymbol{k}}_{g} \Gamma \ket{\Psi (\bm{k})}=c_gu_g
			\Gamma \ket{\Psi (\bm{k})}$. Hence, if $c_g=+1(-1)$, the symmetry eigenvalues
			of $\ket{\Psi (\bm{k})}$ and $\Gamma \ket{\Psi (\bm{k})}$ have the same
			(opposite) signs}\ }\BibitemShut {NoStop}%
\end{thebibliography}

%

\end{document}